\documentclass[runningheads]{llncs}

\usepackage{multicol}
\usepackage{multirow}
\usepackage[table]{xcolor}
\usepackage{graphicx}
\usepackage{longtable}
\usepackage{color}
\usepackage{prettyref}
\newrefformat{fig}{Figure~\ref{#1}}
\newrefformat{tab}{Table~\ref{#1}}
\usepackage{booktabs}
\usepackage{xspace}
\usepackage[disable]{todonotes}
\usepackage{url}
\def\doubleunderline#1{\underline{\underline{#1}}}

\usepackage{framed}
\usepackage{esvect}
\usepackage{tikz}
\usepackage{blkarray}
\usepackage{mathtools}
\usepackage{amsmath}
\usepackage{bm}
\usepackage{adjustbox}
\usepackage{boxedminipage}
\usepackage{blindtext}
\usepackage{multicol}
\usepackage{pgfplots, pgfplotstable}

\usepackage{hyperref}
\usepackage{float}
\usepackage{boxedminipage}
\usepackage{enumitem}
\usepackage{amsmath}
\usepackage{float}
\usepackage[ruled,linesnumbered]{algorithm2e}
\usepackage{subfig}
\usepackage{graphicx}
\usepackage{framed}
\usepackage{esvect}
\usepackage{tikz}
\usepackage{latexsym}

\usepackage{pgfplots, pgfplotstable}
\usepackage{blkarray}
\usepackage{mathtools}
\usepackage{amsmath}
\usepackage{bm}
\usepackage{adjustbox}
\usepackage{blindtext}
\usepackage{multicol}
\usepackage[disable]{todonotes}
\usepackage{prettyref}
\usepackage{multirow}
\usepackage{adjustbox}
\usepackage{blindtext}
\usepackage{multicol}
\usepackage{tablefootnote}
\usepackage{colortbl}
\usepackage[skins]{tcolorbox}
\newtcolorbox{mybox}[2][]{%
  attach boxed title to top center
               = {yshift=-11pt},
  colframe     =black,
  colbacktitle = black,
  title        = #2,#1,
  enhanced,
}

\makeatletter
\DeclareRobustCommand\onedot{\futurelet\@let@token\@onedot}
\def\@onedot{\ifx\@let@token.\else.\null\fi\xspace}

\makeatother

\usepackage{hyperref}
\newcommand{\sss}{\scriptscriptstyle}

\usepackage[primitives,operators,sets,keys,ff,lambda,adversary]{cryptocode}
\newcommand{\nonce}{\ensuremath{{N}}}
\newcommand{\keyt}{\ensuremath{{kt}}}
\newcommand{\counter}{\ensuremath{{ct}}}
\newcommand{\pin}{\ensuremath{\mathsf{PIN}}}
\newcommand{\salt}{\ensuremath{{sa}}}
\newcommand{\trans}{\ensuremath{{t}}}

\renewcommand{\verifier}{\ensuremath{{v}}}
\renewcommand{\state}{\ensuremath{{st}}}
\newcommand{\update}{\ensuremath{\mathsf{Update}}}
\newcommand{\VC}[1]{\ensuremath{#1^{\sss C}}}
\newcommand{\VS}[1]{\ensuremath{#1^{\sss S}}}
\newcommand{\VM}[1]{\ensuremath{#1^{\sss M}}}

\newcommand{\tmp}{\ensuremath{{tmp}}}
\newcommand{\iif}{\ensuremath{\text{If\ }}}
\newcommand{\ssample}{\stackrel{\sss\$}\leftarrow}

\newcommand{\execute}{$\mathtt{Execute}$}
\newcommand{\send}{$\mathtt{Send}$}
\newcommand{\test}{$\mathtt{Test}$}
\newcommand{\reveal}{$\mathtt{Reveal}$}
\newcommand{\corrupt}{$\mathtt{Corrupt}$}
\newcommand{\A}{\mathcal{A}}

\begin{document}
%

\title{A Forward-secure Efficient\\ Two-factor Authentication Protocol}


\author{%
Steven J. Murdoch\thanks{s.murdoch@ucl.ac.uk}\hspace{3mm} and \hspace{3mm} 
Aydin Abadi\thanks{aydin.abadi@ucl.ac.uk}
\institute{University College London (UCL)}}

\maketitle  

\begin{abstract}
Two-factor authentication (2FA) schemes that rely on a combination of knowledge factors (e.g., PIN) and device possession have gained popularity. Some of these schemes remain secure even against strong adversaries that (a) observe the traffic between a client and server, and (b) have physical access to the client’s device, or its PIN, or breach the server.  However, these solutions have several shortcomings; namely, they (i) require a client to remember \emph{multiple} secret values to prove its identity, (ii) involve \emph{several modular exponentiations}, and (iii) are in the \emph{non-standard} random oracle model. In this work, we present a 2FA protocol that resists such a strong adversary while addressing the above shortcomings. Our protocol requires a client to remember only a single secret value/PIN, does not involve any modular exponentiations, and is in a standard model. It is the first one that offers these features without using trusted chipsets. This protocol also imposes up to $40\%$ lower communication overhead than the state-of-the-art solutions do. 

\end{abstract}


\section{Introduction}

The adoption of online services, such as online banking and e-commerce, has been swiftly increasing, and so has the effort of adversaries to gain unauthorised access to such services.  For clients to prove their identity to a remote service provider, they provide a piece of evidence, called an ``authentication factor''. Authentication factors can be based on (i) knowledge factors, e.g., PIN or password, (ii)  possession factors, e.g., access card or physical hardware token, or (iii) inherent factors, e.g., fingerprint.  
Knowledge factors are still the most predominant factors used for authentication \cite{bonneau2010password,JacommeK21}. 
%
%
 The knowledge factors themselves are not strong enough to adequately prevent impersonation 
 %
 %
 \cite{SinigagliaCCZ20,JacommeK21}.  Multi-factor authentication methods that depend on more than one factor are more difficult to compromise than single-factor methods. Recently (on January 26, 2022), the ``Executive Office of the US President'' released a memorandum requiring
the Federal Government's agencies to meet specific cybersecurity standards, including the use of multi-factor authentication, to reinforce the Government’s defences against increasingly sophisticated threat campaigns \cite{Zero-Trust-Cybersecurity}. 
Among multi-factor authentication schemes, two-factor authentication (2FA) methods, including those that rely on a combination of PIN and device possession, have attracted special attention, from banks and e-commerce, due to their low cost and good usability. 

Researchers and companies have proposed various 2FA solutions based on a combination of PIN and device possession. Some of these solutions offer a strong security guarantee against an adversary which may (a) observe the communication between a client and server, and (b) have physical access to the client's device, or its PIN, or breaches the server. These solutions do not rely on trusted chipsets and still ensure that even such a strong adversary cannot succeed during the authentication. Nevertheless, these solutions (i) require a client to remember multiple secret values (instead of a single PIN)  to prove its identity which ultimately harms these solutions' usability, (ii) involve several modular exponentiations that make the device battery power run out fast, and (iii) are in the non-standard random oracle model.

\paragraph{\textbf{\textit{Our Contributions.}}}  In this work, we present a 2FA protocol that resists the strong adversary above while addressing the aforementioned shortcomings and imposing a lower communication cost. Specifically, our protocol:

\begin{itemize}
\item[$\bullet$] requires a client to remember only a single PIN.

\item[$\bullet$] {allows the device to generate a short authentication message.} 

\item[$\bullet$] does not involve any modular exponentiations.

\item[$\bullet$] is in a standard model.

\item[$\bullet$]  imposes up to $40\%$ lower communication costs than the state-of-the-art does.

\end{itemize}

 To attain its goals, our protocol does not use any trusted chipsets; instead, it relies on a novel combination of the following two approaches. First, it requires only the server  (not the device) to verify a client’s PIN. This would allow separating the location where the PIN’s secret key (used to compute the PIN’s authenticator) is stored from the location where the authenticator itself is stored. This approach ensures that an adversary cannot retrieve the PIN, even if it penetrates either location.  Second, it  (a) requires that the server and device use key-evolving symmetric-key encryption (i.e., a combination of forward-secure pseudorandom bit generator and authenticated encryption) to encrypt sensitive messages they exchange,  and (b) requires that used keys be discarded right after their use. This approach ensures the secrecy of the communication between the parties and guarantees that the adversary cannot learn the PIN, even if it eavesdrops on the parties' communication and breaks into the device or server. We formally prove the security of this protocol.

\section{Notations and Preliminaries}
 In this section, we present the main notations and tools used in this work. 
 
 \subsection{Notations}
 
 To disambiguate the different uses of keys and other items of data, variables are annotated with a superscript to indicate their origin. $\VC{\cdot}$ indicates data stored at the client, $\VS{\cdot}$ indicates stored at the server, and $\VM{\cdot}$ indicates the data item has been extracted from a protocol message. A summary of variables can be found in \prettyref{tab:variables}.


\begin{table}[!htb]
\begin{scriptsize}
\footnotesize{
\caption{ \small{Notation table}.}\label{commu-breakdown-party} 
\renewcommand{\arraystretch}{1}
\scalebox{.935}{
\begin{tabular}{p{2cm}@{\hskip 1em} p{5cm}@{\hskip 1em}p{5cm}}

\hline 

\cellcolor{gray!30} \scriptsize \textbf{Symbol}&\cellcolor{gray!30} \scriptsize\textbf{Purpose}&\cellcolor{gray!30} \scriptsize\textbf{Source and lifetime}  \\
    \hline
    
     \hline

\cellcolor{white!20}\scriptsize$\prf(.)$ &\cellcolor{white!20}\scriptsize  Pseudorandom function.&\cellcolor{white!20}\scriptsize Used to derive a verifier and session key.\\

\cellcolor{gray!20}\scriptsize FS-PRG &\cellcolor{gray!20}\scriptsize  Forward-secure Pseudorandom Bit Generator.&\cellcolor{gray!20}\scriptsize Used to derive temporary keys.\\

 \cellcolor{white!20}\scriptsize \VC{k}, \VS{k} &\cellcolor{white!20}\scriptsize Authenticated Encryption (AE) key at the client and server sides respectively.&\cellcolor{white!20}\scriptsize {Key $k$ randomly generated by the system operator and stored by the client as \VC{k} and server as \VS{k} at device creation. Constant for the lifetime of the device.}\\   

  \cellcolor{gray!20}\scriptsize \VC{\state}, \VS{\state}&\cellcolor{gray!20}\scriptsize {The state of FS-PRG at the client and server sides respectively.}&  \cellcolor{gray!20}\scriptsize Initialised to randomly generated state $\state_{\sss 0}$ at device creation. Updated using FS-PRG. \\   

  \cellcolor{white!20}\scriptsize  \VC{\keyt_{\sss 1}}, \VS{\keyt_{\sss 1}} &\cellcolor{white!20}\scriptsize Temporary key for the enrolment phase.& \cellcolor{white!20}\scriptsize Output by FS-PRG and used for a single message exchange before being discarded.\\   
  \cellcolor{gray!20}\scriptsize \VC{\keyt_{\sss 2}}, \VS{\keyt_{\sss 2}}, \VC{\keyt_{\sss 3}}, \VS{\keyt_{\sss 3}} &\cellcolor{gray!20}\scriptsize Temporary keys of $\mathtt{PRF}$, used in the authentication phase.&\cellcolor{gray!20}\scriptsize Output by FS-PRG and used for a single message exchange before being discarded. \\   

    \cellcolor{white!20}\scriptsize \VC{\counter}, \VS{\counter} &\cellcolor{white!20}\scriptsize Counter for synchronising FS-PRG state and detecting replayed messages.& \cellcolor{white!20} \scriptsize Initialised to zero at device creation. \VC{\counter} and \VS{\counter} are updated atomically along with \VC{\state} and \VS{\state} respectively. \\
   \cellcolor{gray!20}\scriptsize   \VS{\nonce}, \VM{\nonce} & \cellcolor{gray!20}\scriptsize Random challenge for detecting replayed messages.& \cellcolor{gray!20}\scriptsize Generated randomly by the server for each message.\\
\cellcolor{white!20}\scriptsize  \VC{\salt}&\cellcolor{white!20}\scriptsize   Random PIN-obfuscation secret key. &\cellcolor{white!20}\scriptsize Initialised to randomly generated value at device creation. Not known by the server or system operator. \\ 
\cellcolor{gray!20}\scriptsize \VC{\pin} &\cellcolor{gray!20}\scriptsize  Client's PIN. &\cellcolor{gray!20}\scriptsize Entered by the client. It is never stored in the device and used to generate a verifier. \\      
\cellcolor{white!20}\scriptsize \VC{\verifier}, \VS{\verifier}, \VM{\verifier} &\cellcolor{white!20}\scriptsize  Verifier, generated from PIN-obfuscation key and the client's PIN. &\cellcolor{white!20}\scriptsize  Stored by the server after the enrolment phase. It is not stored by the client. \\  

 \cellcolor{gray!20}\scriptsize  \VS{\trans}, \VM{\trans} & \cellcolor{gray!20}\scriptsize Description of a transaction to be authenticated. & \cellcolor{gray!20}\scriptsize  Generated and sent by the server.\\
\cellcolor{white!20}\scriptsize \VC{response} &\cellcolor{white!20}\scriptsize Authentication response. &\cellcolor{white!20}\scriptsize Computed by the client.\\   

\cellcolor{gray!20}\scriptsize  \VS{expected} &\cellcolor{gray!20}\scriptsize Expected authentication response. &\cellcolor{gray!20}\scriptsize Computed by the server.\\

 \hline

 \hline

\end{tabular}\label{tab:variables}
}
}
\end{scriptsize}
\end{table}


\subsection{Pseudorandom Function}\label{subsec:PRF}

Informally, a pseudorandom function (\prf) is a deterministic function that takes as input a key and some argument. It outputs a value  indistinguishable from that of a truly random function with the same domain and range.   A $\prf$ is formally defined as follows \cite{KatzLindell2014}. 
\begin{definition} Let $\prf:\{0,1\}^{\sss\psi}\times \{0,1\}^{\sss \eta}\rightarrow \{0,1\}^{\sss  \lambda}$ be an efficient  keyed function. It is said $\prf$ is a pseudorandom function if for all probabilistic polynomial-time distinguishers $B$, there is a negligible function, $\mu(.)$, such that:

\begin{equation*}
\bigg | \Pr[B^{\sss \prf_{\hat{k}}(.)}(1^{\sss \psi})=1]- \Pr[B^{\sss \omega(.)}(1^{\sss \psi})=1] \bigg |\leq\mu(\psi)
\end{equation*}
where  the key, $\hat{k}\stackrel{\sss\$}\leftarrow\{0,1\}^{\sss\psi}$, is chosen uniformly at random and $\omega$ is chosen uniformly at random from the set of functions mapping $\eta$-bit strings to $\iota$-bit strings. We define $Adv^{\sss\prf}(\adv)$ as the advantage of the adversary which interacts with pseudorandom and random functions. 

\end{definition}

Since a pseudorandom function is deterministic and outputs the same value if queried twice on the same inputs, when proving a protocol that uses a $\prf$, it is assumed that the distinguisher never queries oracles $\prf$ and $\omega$ twice on the same inputs \cite{KatzLindell2014}.

%

\subsection{Forward-Secure Pseudorandom Bit Generator}
A Forward-Secure Pseudorandom Bit Generator (FS-RPG), is a \emph{stateful} object which consists a pair of algorithms and a pair of positive integers, i.e., $\mathsf{FS\text{-}RPG} = \Big((\mathsf{FS\text{-}RPG}.\kgen, $ $ \mathsf{FS\text{-}RPG.next}),( b, n)\Big)$, as defined in \cite{BellareY03}.  The probabilistic key generation algorithm $\mathsf{FS\text{-}RPG}.\kgen$ takes a security parameter as input and outputs an initial state ${st}_{\sss 0}$ of length $s$ bits. $\mathsf{FS\text{-}RPG.next}$ is a key-updating algorithm which, given the current state ${st}_{\sss i-1}$, outputs a pair of a $b$-bit block ${out}_{\sss i}$ and the next state ${st}_{\sss i}$. We can produce a sequence  ${out}_{\sss 1},..., {out}_{\sss n}$  of  $b$-bit output blocks, by first generating a key  ${st}_{\sss 0}\stackrel{\sss \$}\leftarrow\mathsf{FS\text{-}RPG}.\kgen(1^{\sss \lambda})$ and then running $({out}_{\sss i}, st_{\sss i})\leftarrow  \mathsf{FS\text{-}RPG.next} (st_{\sss i-1})$ for all $i, 1\leq i\leq n$. As with a standard pseudorandom bit generator, output blocks of this generator should be computationally indistinguishable from a random bit
string of the same length. The additional property required from a
FS-RPG is that even when the
adversary learns the state, output blocks generated before the point of
compromise remain computationally indistinguishable from random bits.
This requirement implies that it is computationally infeasible to
recover a previous state from the current state. We restate a formal definition and construction of a forward-secure pseudorandom bit generator in Appendix \ref{sec::def-FS-PRG}.

Recall, $\mathsf{FS\text{-}RPG.next}$ updates the state of the forward-secure generation by one step; however, our protocol sometimes needs to invoke $\mathsf{FS\text{-}RPG.next}$ multiple times sequentially. Thus, for the sake of simplicity, we define a wrapper algorithm $ \update(\state_{\sss a}, d)$ which wraps  $\mathsf{FS\text{-}RPG.next}$. Algorithm $\update$  as input takes a current state (similar to $\mathsf{FS\text{-}RPG.next}$) and new parameter $d$ that determines how many times   $\mathsf{FS\text{-}RPG.next}$ must be invoked internally. It invokes  $\mathsf{FS\text{-}RPG.next}$ $d$ times and outputs the pair  $({out}_{\sss b}, \state_{\sss b})$  which are the output of $\mathsf{FS\text{-}RPG.next}$ when it is invoked for $d$-th time, where $ b> a$.


 \vspace{2mm}
 
\subsection{Authenticated Symmetric-key Encryption} 
Informally, authenticated encryption $\Pi=(\mathsf{Gen}, \mathsf{Enc}, \mathsf{Dec})$ is an encryption scheme that simultaneously ensures the secrecy and integrity of a massage. It can be built via symmetric or asymmetric key encryptions. In this work, we use authenticated symmetric-key encryption, due to its efficiency. $\mathsf{Gen}$ is a probabilistic key generating algorithm that takes a security parameter and returns an encryption key $k$. $\mathsf{Enc}$ is a deterministic encryption algorithm that takes the secret key $k$ and a message $m$, it returns a ciphertext $M$ along with the corresponding tag $t$. $\mathsf{Dec}$ is a deterministic algorithm that takes the ciphertext $M$, the tag $t$, and the secret key $k$. It first checks the tag's validity, if it accepts the tag, then it decrypts the message and returns $(m,1)$. Otherwise, it returns $(.,0)$.

The security of such encryption consists of the notion of secrecy and integrity. The secrecy notion requires that the encryption be secure against chosen-ciphertext attacks, i.e., CCA-secure. The notion of integrity considers existential unforgeability under an adaptive chosen message attack. We refer readers to \cite{KatzLindell2014} for a formal definition of authenticated symmetric-key encryption.




\vspace{-2mm}
\section{Threat Model}\label{sec::model}

In this section, we present the threat model that we consider in this work. A two-factor authentication scheme involves two players:

\begin{itemize}
\item[$\bullet$] {Client ($C$)}: It is an honest party which tries to prove its identity to a server by using a combination of a PIN and a device.

\


\item[$\bullet$] {Server ($S$)}:  It is a semi-honest (or passive) adversary which follows the protocol's instruction and tries to learn $C$'s PIN. It also tries to authenticate itself to $C$.  
\end{itemize}

We allow the server to communicate with the device through the client. In particular, similar to various previous works (e.g., those in \cite{JareckiJKSS21,Digipass-website,Gemalto}), we assume the device has a camera that lets the device scan the (QR code) messages the server sends to it via the client. Each of the above parties may have several instances possibly running concurrently. In this work, we denote instances of client and server by  $C^{i}$ and  $S^{j}$ respectively. Each instance is called an oracle.  
 %
%
To formally capture the capabilities of an adversary $\mathcal{A}$ in a hardware token-based 2FA,  we mainly use the (adjusted) model proposed by Bellare \textit{et al.} \cite{BellarePR00}. In this model, the adversary’s capabilities are cast via various queries that it sends to different oracles, i.e., instances of the honest parties; the client and server interact with each other for some fixed number of flows, until both instances have terminated. By that time, each instance should have accepted holding a particular session key ($sk$), session id ($SID$), and partner id ($PID$). At any point in time, an oracle may ``\emph{accept}''. When an oracle accepts, it holds $sk$, $SID$, and $PID$. A client instance and a server instance can accept at most once. The above model was initially proposed for the password-based key exchange schemes in which the adversary does not corrupt either player.  Later, Wang \textit{et al.} \cite{WangW18}  added a few more queries to the model of  Bellare \textit{et al.} to make it suitable for two-factor authentication schemes. The added queries would allow an adversary to learn either of the client's factors (i.e., either PIN or secret parameters stored in the hardware token) or the server's secret parameters.  Below, we restate the related queries.

\begin{itemize}
\item [$\bullet$] \execute($C^{i}, S^{j}$): this query captures passive attacks in which the adversary, $\mathcal{A}$, has access to the messages exchanged between $C^{i}$ and $S^{j}$ during the correct executions of a 2FA protocol, $\pi$. 

\

\item [$\bullet$] \reveal($I$): this query models the misuse of the session key $sk$ by instance $I$.  Adversary $\mathcal{A}$ can use this query if $I$ holds a session key; in this case, upon receiving this query, $sk$ is given to $\A$. 

\

\item [$\bullet$] \test($I$): this query models the semantic security of the session key. It is sent at most once by $\A$ if the attacked instance $I$  is ``fresh'' (i.e., in the current protocol execution $I$ has accepted and neither it nor the other instance with the same $SID$ was asked for a Reveal query). This query is answered as follows. Upon receiving the query, a coin $b$ is flipped. If $b=1$, then session key $sk$ is given to $\A$; otherwise (if $b=0$), a random value is given to $\A$. 

\

\item [$\bullet$] \send($I, m$):  this query models active attacks where $\mathcal{A}$ sends a message, $m$, to instance $I$ which follows  $\pi$'s instruction, generates a response, and sends the response back to $\mathcal{A}$.  Query  \send($C^{\sss i}, \text{start}$) initialises $\pi$; when it is sent, $\mathcal{A}$ would receive the message that the client would send to the server. 

%

\

\item [$\bullet$]  \corrupt($I, a$): this query models the adversary's capability to corrupt the involved parties.

\begin{itemize}

\item if $I=C$:  it can learn (only) one of the factors of $C$. Specifically, 

\

\begin{itemize}
\item if $a=1$, it outputs $C$'s PIN. 

\

\item if  $a=2$, it outputs all parameters stored in the hardware token. 
  \end{itemize}
  
  \
  
\item if $I=S$, it outputs all parameters stored in $S$. 
\end{itemize}
\end{itemize}

\subsubsection{Authenticated Key Exchange (AKE)  Security.} Security notions (i.e., session key's semantic security and authentication) are defined with regard to the executing of protocol $\pi$, in the presence of   $\mathcal{A}$. To this end, a game $Game^{\sss ake} (\mathcal{A},\pi)$ is initialized by drawing a PIN  from the PIN's universe,
providing coin tosses to $\A$ as well as to the oracles, and then running the adversary by letting it ask a polynomial
number of queries defined above. At the end of the game, $\A$  outputs its guess $b'$ for  bit $b$
involved in the Test-query. 

\

\noindent\textit{Semantic security.} It requires that the privacy of a session key be preserved in the presence of $\A$, which has access to the above queries. We say that $\A$  wins if it manages to correctly guess bit $b$ in the Test-query, i.e., manages to output $b'=b$. We denote its advantage as the probability that $\A$  can correctly guess the value of $b$; specifically, such an advantage is defined as $Adv_{\sss\pi}^{\sss ss}(\A)=2Pr[b=b']-1$, where the probability space is over all the random coins of the adversary and all the oracles\footnote{Note, more accurate advantage is $Adv_{\sss\pi}^{\sss ss}(\A)=Pr[b=b']-\frac{1}{2}$; however,  for simplicity in \cite{BellarePR00}, the righthand side of the equation was multiplied by $2$. Since we are in the security context, the advantage must still remain small. We use the same convention.}.  Protocol $\pi$ is said to be semantically secure if $\A$'s advantage is negligible in the security parameter, i.e., $Adv_{\sss\pi}^{\sss ss}(\A)\leq \mu(\lambda)$.


\

\noindent\textit{Authentication.} It requires that $\A$ must not be able: (a)  to impersonate $C$, even if it has access to the traffic between the two parties as well as having access to either $C$'s PIN, or its authentication device, or (b) to impersonate  $S$, although it has access to the traffic between the two parties.   We say that   $\A$  violates mutual authentication if some oracle accepts a session key and terminates, but has no partner oracle, which shares the same key.  Protocol $\pi$ is said to achieve mutual authentication if for any adversary $\A$  interacting with the parties, there exists a negligible function $\mu(.)$ such that for any security parameter $\lambda$ the advantage of $\A$ (i.e., the probability of successfully impersonating a party)  is negligible in the security parameter, i.e., 
$Adv_{\sss\pi}^{\sss aut}(\mathcal{A})\leq \mu(\lambda)$.

In certain schemes (including ours), during the key agreement and authentication phase, the client needs to also verify a message, e.g., a bank transaction. To allow such verification to be carried out deterministically, which will be particularly useful in the schemes proof, we define a predicate $y\leftarrow \phi(m, \pi)$, where $y\in \{0,1\}$. This predicate takes as input a message $m$ (e.g., bank's transactions) and a policy $\pi$ (e.g., a client's policy specifying a payment amount and destination account number). It checks if the message matches the policy. If they match, it outputs $1$; otherwise, it outputs $0$.




\section{{Straw-man Solutions}}
In this section, we provide an overview of a couple of solutions that seem to work and discuss their shortcomings. 

\subsection{Straw-man Proposal I}\label{sec:straw-man-1}

\begin{figure}[!thb]
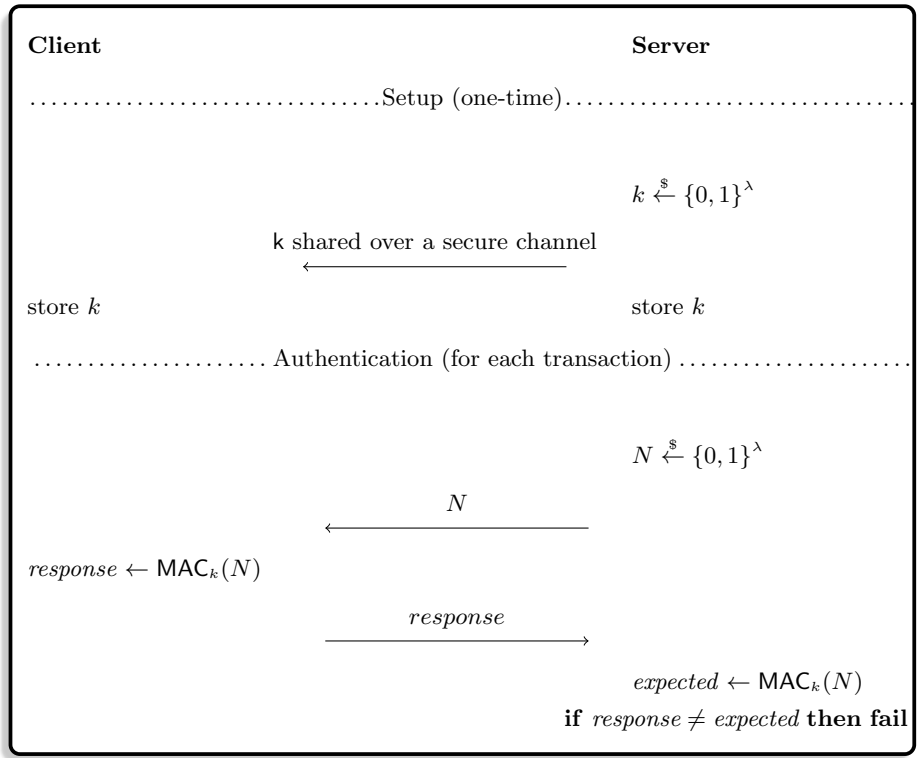

\begin{tcolorbox}[enhanced,width=4.75in, height=100mm, left=1mm,
    drop fuzzy shadow southwest,
    colframe=black,colback=white]
    {\small{

 \procedure{}{%
 \textbf{Client} \<  \textbf{Server} \pclb
  \pcintertext[dotted]{Setup (one-time)} \\ 
    \<  k \stackrel{\sss \$}\leftarrow \{0,1\}^{\sss \lambda} \pclb
  \< \sendmessageleft*{\text{\key{} shared over a secure channel}} \< \\
  \text{store}\ k \<  \text{store}\ k \pclb
  \pcintertext[dotted]{Authentication (for each transaction)} \\
  \<  \nonce{} \stackrel{\sss \$}\leftarrow \bin^{\sss\secpar} \pclb
  \< \sendmessageleft*{\nonce} \< \\
  \text{\textit{response}} \gets \mac_{\sss k}(\nonce) \<  \pclb
  \< \sendmessageright*{response} \< \\
  \<  \text{\textit{expected}} \gets \mac_{\sss k}(\nonce) \\
  \<\hspace{-9mm}    \pcif \textit{response} \ne \textit{expected} \pcthen \pcfail
   }
    \caption{Straw-man protocol I.}
    
    \label{fig:strawman1}
    }}
    \end{tcolorbox}
\end{figure}

A simple authentication protocol would be for the server to generate a secret key $k$, then enrol a client's device by sharing this key over a secure channel, e.g., loaded onto the device at the time of manufacture.
Then, when the secure channel is not available (e.g., during Internet banking) the server sends a
randomly generated challenge to the client which replies with a Message Authentication Code (MAC)  computed of this challenge, under $k$.
This protocol provides the server assurance
that the response originated from the correct client and
bounds the time at which the response was generated to be between the time that the challenge was sent and when the response was received. 
%
%
The resulting protocol is shown in \prettyref{fig:strawman1}.  An alternative design would be to omit the challenge message containing
the nonce. For example, we could compute the MAC of a counter. However, this protocol is vulnerable to a pre-play attack, where a
response is collected and replayed at a later time. Alternatively, the
MAC could be computed of a timestamp. This approach
gives the server assurance of when the response was generated.  But, it
requires the device to have a real-time clock. Doing so
would increase power requirements and limit the device's lifespan because
the battery could not be replaced by the client without desynchronising
the clock. Alternatively, a backup battery could be included, but this
would significantly increase the device cost.

All of these protocols have a major weakness, if the device is stolen, then the adversary can generate valid
authentication responses. 

\subsection{Straw-man Proposal II}\label{sec:straw-man-2}

We extend the previous challenge-response protocol to compute the response
over both the challenge and a client's PIN, to prevent an adversary who
has stolen the device from completing the authentication phase. This creates a
2FA scheme, depending on something the client has (i.e., the
authentication device) and something the client knows (i.e., the PIN). The
server can compute the
expected response and validate the response produced by the device. If this validation succeeds, then the server has the
same assurances of Straw-man Proposal 1 and additionally knows that the correct
PIN was entered into the device. To this end, we could store the PIN on the server. 

\begin{figure}[!htb]
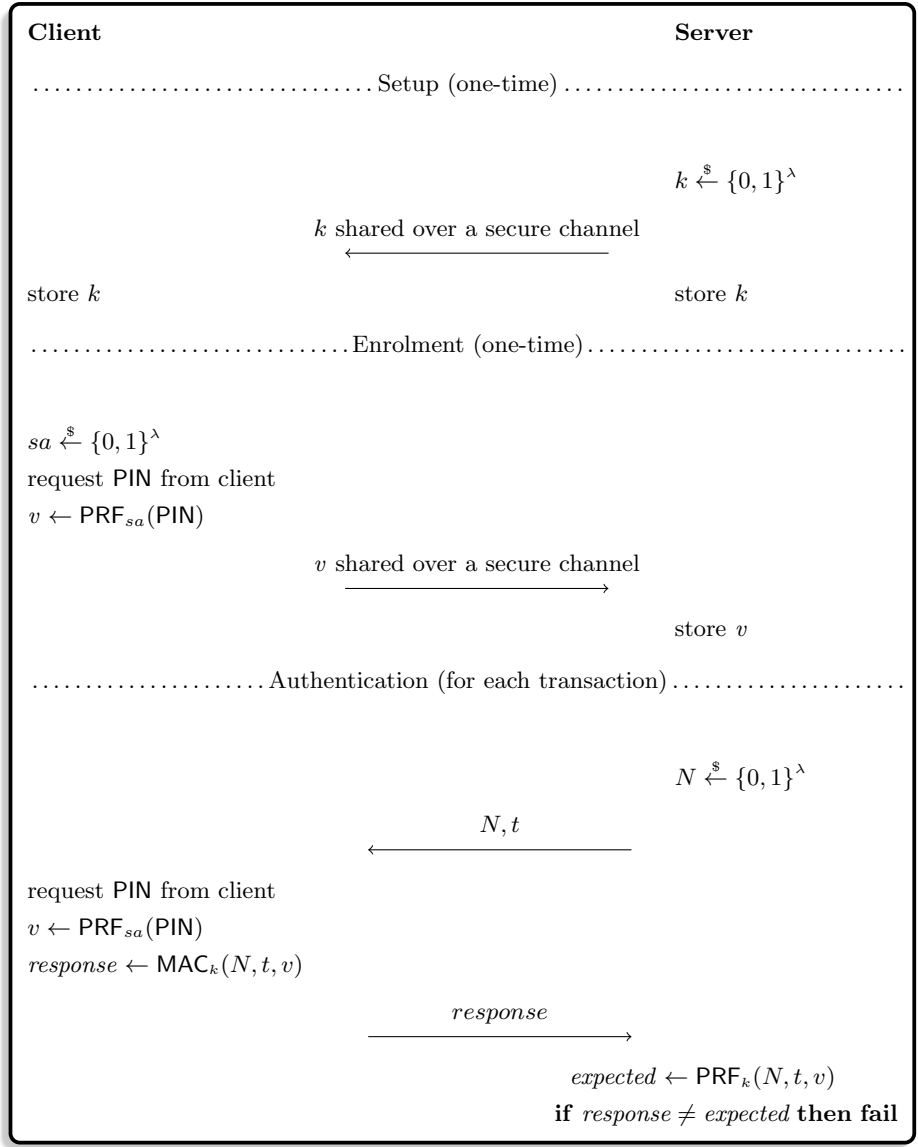

\begin{tcolorbox}[enhanced,width=4.75in, height=152mm, left=1mm,top=0.5mm,
    drop fuzzy shadow southwest,
    colframe=black,colback=white]
 \procedure{}{%
 \textbf{Client} \< \textbf{Server} \pclb
  \pcintertext[dotted]{Setup (one-time)} \\
  \<   k \stackrel{\sss \$}\leftarrow \{0,1\}^{\sss \lambda} \pclb
  \< \sendmessageleft*{k\text{\ shared over a secure channel}} \< \\
  \text{store}\ k \<  \text{store}\ k \pclb
  \pcintertext[dotted]{Enrolment (one-time)} \\ 
  \salt \stackrel{\sss \$}\leftarrow \{0,1\}^{\sss \lambda} \< \< \\
  \text{request \pin{} from client} \< \< \\
  \text{\textit{v}} \gets \prf_\salt(\pin) \< \< \pclb
  \< \sendmessageright*{\text{\textit{v} shared over a secure channel}} \< \\
  \<  \text{store \textit{v}} \pclb
  \pcintertext[dotted]{Authentication (for each transaction)} \\
  \<  \nonce{} \stackrel{\sss \$}\leftarrow \bin^{\sss\secpar} \pclb
  \< \sendmessageleft*{\nonce, t} \< \\
  \text{request \pin{} from client} \< \< \\
  v \gets \prf_\salt(\pin) \< \< \\
  \text{\textit{response}} \gets \mac_{\sss k}(\nonce, t, v) \< \< \pclb
  \< \sendmessageright*{response} \< \\
  \<\hspace{-14mm}  \text{\textit{expected}} \gets \prf_{\sss k}(\nonce, t, v) \\
  \< \hspace{-16mm} \pcif \textit{response} \ne \textit{expected} \pcthen \pcfail
   }
    \caption{Straw-man protocol II.}
    \label{fig:strawman2}
    
    \end{tcolorbox}
\end{figure}

Nevertheless, it is undesirable for the server to know the PIN, as the client may
use the same PIN for other unrelated purposes. Having the server store
the hash of the PIN would not help because the low entropy of a
convenient PIN (around 13 bit for a 4-digits) is trivially vulnerable to
a brute-force pre-image attack. We can avoid this problem by replacing
the PIN in the protocol with a verifier, which is the output of a PRF computed over the PIN
under a secret key held only by the device. This
verifier must be sent to the server when the device is enrolled. Given
the correct PIN, the device could compute the verifier. In this case, a corrupt server
would not be able to recover the PIN from the verifier, without
knowledge of the secret key.
 We can incorporate a description of the transaction into the challenge
message and computation of the authentication response. This transaction
is generated by the server to indicate to the client what action will be
performed if the authentication succeeds. The resulting protocol is shown in 
\prettyref{fig:strawman2}.

An alternative protocol design would be to store the PIN on the device,
and for the device to only permit the authentication key $k$ to be used if
the PIN is entered correctly. For this design to be
resistant to an adversary who has stolen the device, it must not be feasible to extract the PIN and  must not be feasible to bypass the PIN verification. This functionality requires security features not commonly
available on low-cost microcontrollers. Security assured co-processors
are available, but would substantially increase the cost of the device. 
The protocol above meets many desirable criteria for an authentication
protocol. Specifically, a verified authentication response gives the server assurance that (a)
 the device is present, due to the random nonce, (b)
 the correct device was used, due to the use of the key, (c) the device has
not been stolen, due to the PIN, and (d) the client saw the transaction that
the server is about to perform, due to the inclusion of the transaction's description within the MAC computation. The protocol does not require
the device to have a real-time clock, so a single client-replaceable battery may be used. The PIN is also not stored by the device and so no
special tamper-resistant hardware is necessary. There is no need to
protect the authentication key against physical tampering because anyone
with access to the device could simply use the device to perform
authentication.

There is still a remaining serious risk. Let us suppose that the adversary has
recorded a valid authentication response and the corresponding
challenge. The adversary who has access to the device can extract the authentication key. Now, the
adversary has all the information needed to locally brute-force the short PIN, and then generate a valid response to any future authentication challenge
from the server. 

\vspace{60mm}

\section{The Protocol}\label{sec::the-protocol}

In this section, we present an efficient 2FA protocol that remains secure even if an adversary (a) observes the traffic between a client and server, and (b) has access to the client’s device, or its PIN, or breaches the server. To design a protocol that can offer the above features, we rely on a novel combination of the following two approaches. First, we require the client's  PIN verification to take place only on the server. This allows us to separate the location where the PIN's secret key (used to generate the PIN's authenticator) is stored from the location where the authenticator is stored. This approach ensures that even if either location is breached, then the adversary would not have sufficient information to retrieve the PIN even through brute-forcing all possible PINs. Our observation is that even in this setting, the device can perform a basic check to detect the client's mistake without having to permanently store the (representation of the) PIN; for instance, this can be done by asking the client to type in the PIN twice and checking if the two entries match with each other.

Second, we (a) require every sensitive message exchanged between the server and client to be encrypted using key-evolving symmetric-key encryption (i.e., a combination of forward-secure pseudorandom bit generator and authenticated encryption) and (b) require the used keys, of key-evolving symmetric-key encryption, to be discarded right after their use. This approach ensures the secrecy of the communication between the parties and ensures that if the device or server is broken in, the adversary cannot learn the past communication to learn the PIN, with the assistance of the information it extracts from the breached location.

Our protocol consists of three main phases; namely, (i) a setup phase, performed once when the authentication device is manufactured, (ii) an enrolment phase for setting or changing a client's PIN, and (iii) an authentication phase in which the actual authentication is performed. As we already stated, each party has a unique (public) ID. In the protocol, we assume the parties include their IDs in their outgoing messages.  Similar to other (two-factor) authentication schemes, we assume the server maintains a local threshold, and if the number of incorrect responses from a client within a fixed time exceeds the threshold, then the client and its device will be locked out. Such a check is implicit in the protocol's description. In the remaining of this section, we describe each phase.

\subsection{Setup Phase}
\label{sec:setup}

\begin{figure}[!htbp]
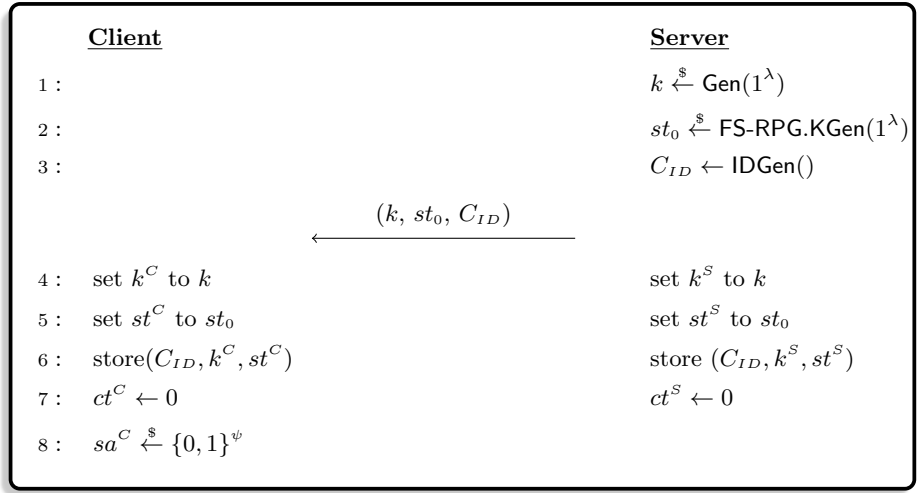

\begin{center}
    \begin{tcolorbox}[enhanced,width=4.75in, height=65mm, left=1mm,top=0.1mm,
    drop fuzzy shadow southwest,
    colframe=black,colback=white]
 \centering
 \procedure{}{%
 \hspace{8mm}\underline{\textbf{Client}} \< \<\hspace{4mm} \underline{\textbf{Server}} \\
 \pcln \< \<\hspace{4mm} k \ssample \mathsf{Gen}(\secparam) \\
 \pcln \< \<\hspace{4mm} \state_{\sss 0} \ssample \mathsf{FS\text{-}RPG}.\kgen(\secparam) \\
  \pcln \< \<\hspace{4mm} C_{\sss ID} \gets \mathsf{IDGen}() \pclb
 \< \sendmessageleft*{(\text{$k$,  $\state_{\sss 0}$, $C_{\sss ID}$)}} \< \\
 \pcln \text{set  $\VC{k}$ to $k$} \< \<\hspace{4mm} \text{set $\VS{k}$ to $k$}\\
 \pcln \text{set $\VC{\state}$ to  $\state_{\sss 0}$} \< \<\hspace{4mm} \text{set  $\VS{\state}$ to $\state_{\sss 0}$}\\
  \pcln \text{store$(C_{\sss ID}, \VC{k}, \VC{\state})$} \< \<\hspace{4mm} \text{store $(C_{\sss ID}, \VS{k}, \VS{\state})$}\\
 \pcln \VC{\counter} \gets 0 \< \<\hspace{4mm} \VS{\counter} \gets 0 \\
 \pcln \VC{\salt} \ssample \{0,1\}^{\sss \psi} \< \<\hspace{4mm} \\
 }
\end{tcolorbox}
\end{center}
    \caption{Setup phase.}
    \label{fig:setup}
\end{figure}

To bootstrap the protocol, in the setup phase, we require that the client
and server share \emph{initial} randomly generated key $k$ for AE and key $\state_{\sss 0}$ for  FS-PRG.
The counter for the FS-PRG state is set to $0$ on both sides. 
These values could be securely loaded into the device at the time of
manufacture or can be sent (via a secure channel) to the client who can use the device camera to scan and store them in the device. In this phase, the device generates and locally stores a random secret key $\VC{\salt}$ for \prf.  \prettyref{fig:setup} presents the setup in detail.

\subsection{Enrolment Phase}
\label{sec:enrollment}

\begin{figure}[!htb]
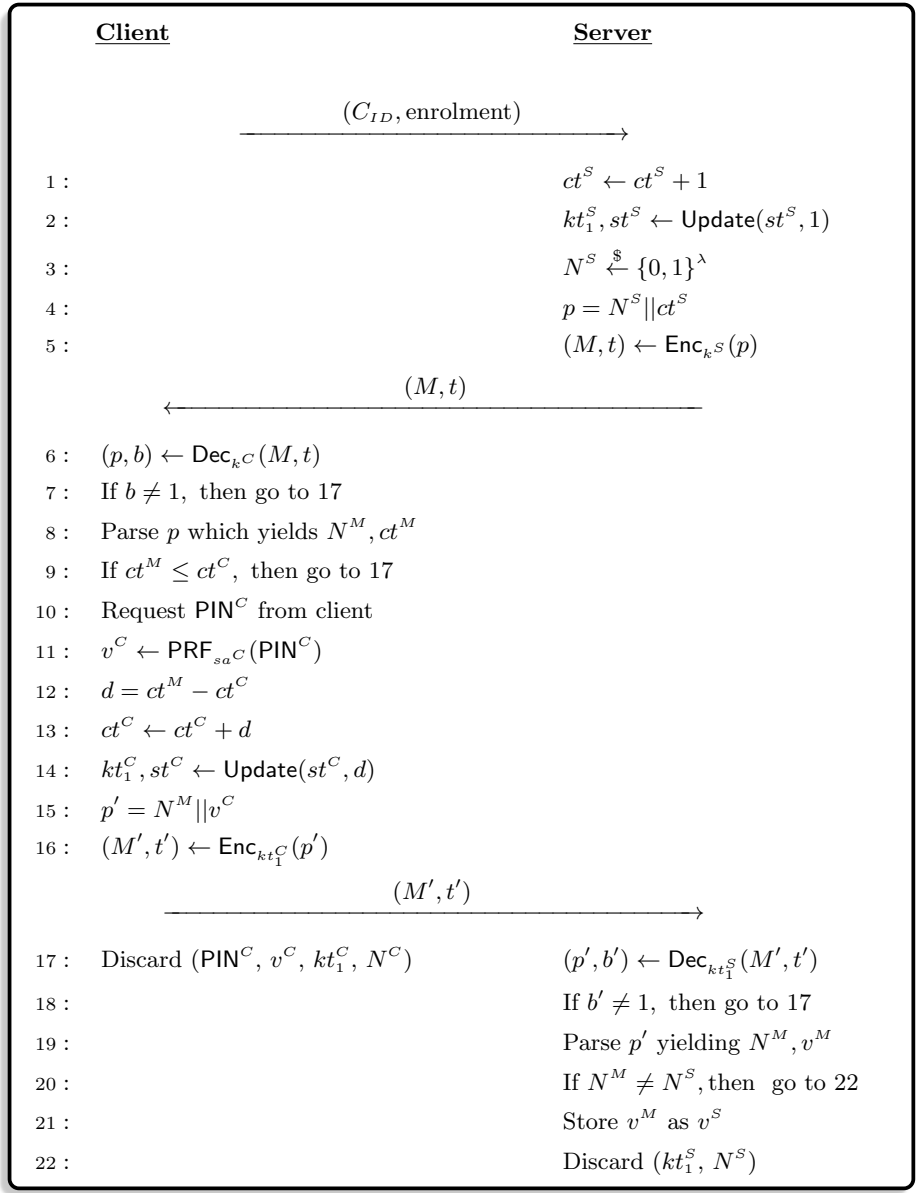

\setlength{\fboxsep}{1pt}
\begin{center}
    \    \begin{tcolorbox}[enhanced,width=4.75in, height=158mm, left=1mm,top=.5mm,
    drop fuzzy shadow southwest,
    colframe=black,colback=white]
 \centering
 \procedure{}{%
  \hspace{8mm}\underline{\textbf{Client}} \< \hspace{12ex} \<\hspace{4mm} \underline{\textbf{Server}} \\
 %
\pclb
 \sendmessagerightx[5cm]{6}{(C_{\sss ID},  \text{enrolment})} \\
 \pcln \< \<\hspace{2.6mm} \VS{\counter} \gets \VS{\counter} + 1  \\
 \pcln \< \<\hspace{2.6mm} \VS{\keyt_{\sss 1}}, \VS{\state} \gets \update(\VS{\state}, 1)\\
 %
 %
 \pcln \< \<\hspace{2.6mm} \VS{\nonce} \stackrel{\$}\leftarrow \bin^{\sss\secpar}  \\
 \pcln \< \<\hspace{2.6mm} p= \VS{\nonce}|| \VS{\counter}\\ 
 \pcln \< \<\hspace{2.6mm} {(M,t) \gets \enc_{\sss\VS{k}}(p) } \pclb
 \sendmessageleftx[7cm]{6}{(M,t) } \\
 %
  \pcln (p,b) \gets \mathsf{Dec}_{\sss\VC{k}}(M,t) \< \< \\
  \pcln \iif b\neq 1,    \text{\ then\ go\ to\ } \ref{enroll:fail} \\ 
   \pcln \text{Parse\ } p \text{\ which\ yields\ } \VM{\nonce}, \VM{\counter} \< \< \\ 
 \pcln\label{enroll:clinet-check-counter}  \iif \VM{\counter} \le \VC{\counter}, \text{\ then\ go\ to\ } \ref{enroll:fail} \\ 
 %
 \pcln \text{Request \VC{\pin} from client} \< \< \\
 \pcln \VC{\verifier} \gets \prf_{\sss\VC{\salt}}(\VC{\pin}) \< \< \\
 %
 \pcln d=  \VM{\counter}-\VC{\counter} \< \< \\
 %
  %
   \pcln   \VC{\counter} \gets \VC{\counter} + d\< \< \\
    \pcln \VC{\keyt_{\sss 1}},  \VC{\state} \gets \update(\VC{\state}, d) \< \< \\ 
 %
 %
 \pcln p'=  \VM{\nonce} || \VC{\verifier}\< \< \\ 
 \pcln (M',t') \gets \enc_{\sss\VC{\keyt_{\sss 1}}}(p')\< \< \pclb
 \sendmessagerightx[7cm]{6}{(M',t') } \\
  \pcln \label{enroll:fail} \text{Discard (\VC{\pin}, \VC{\verifier}, \VC{\keyt_{\sss 1}},  \VC{\nonce}})
 \< \<\hspace{2.6mm} (p',b') \gets \mathsf{Dec}_{\sss\VS{\keyt_{\sss 1}}}(M',t') \< \< \\
   \pcln\< \<\hspace{2.6mm} \iif b'\neq 1,    \text{\ then\ go\ to\ } \ref{enroll:fail} \\ 
 \pcln\< \<\hspace{2.6mm} \text{Parse\ } p' \text{\ yielding\ }  \VM{\nonce},  \VM{\verifier} \< \< \\
 \pcln \< \<\hspace{2.6mm} \iif \VM{\nonce} \ne \VS{\nonce}, \text{then \ go\ to\ } \ref{enroll:fail-server-} \\
 \pcln \< \<\hspace{2.6mm} \text{Store \VM{\verifier} as \VS{\verifier}} \\
 %
 %
  \pcln \label{enroll:fail-server-} \< \<\hspace{2.6mm}  \text{Discard (\VS{\keyt_{\sss1}}, \VS{\nonce}}) \\ 
 }
\end{tcolorbox}
\end{center}
    \caption{Enrolment phase.}
    \label{fig:enrollment}
\end{figure}

The goal of the enrolment phase is to set the client's PIN, without providing the  
 server with sufficient information to discover this PIN. The server allows this phase to take place only over a channel through which the client has already proven their identity.  At the end of this phase, the server will have stored the verifier $\verifier$ corresponding to the client's selected PIN.
The steps involved in this phase are shown in \prettyref{fig:enrollment} in detail.

We briefly explain how this phase works.  The server first updates the FS-PRG's state, which results in a new state and random value \VS{\keyt_{\sss 1}}; it also increments its counter by one. Then, the server generates a random challenge \VS{\nonce}. The server sends the enrolment challenge message which is a combination of the current counter and the challenge encrypted via the AE under the shared key $k$.
 On receiving this message, the client uses its device to scan the (QR) message it receives. The device decrypts the message using $k$ that was shared with the server during the setup phase.  If decryption succeeds, it  extracts the server's challenge and counter from the message.  If the device's local counter is greater than the counter it received from the server, it would be impossible for the device to recover the $\VS{\keyt_{\sss 1}}$ that the server will use, so the protocol must abort here. As we will discuss in \prettyref{sec:synchronisation}, this case would not occur, with a high probability. Next, the device requests the PIN from the client and ensures it is what the client intends by for example requesting it twice and checking that match.

The device then generates a verifier $\VC{\verifier}$, by deriving a pseudorandom value from the PIN using $\prf$ and the random key $\VC{\salt}$ it generated in the setup phase.  After that, the device locally synchronises the FS-PRG's state with the server by updating the state until it matches the counter received from the server; this yields \VC{\keyt_{\sss 1}}. This synchronisation is possible because the check at line \ref{enroll:clinet-check-counter} has already assured the device that its state is behind the server's state by at least one step. After the update, \VC{\keyt_{\sss 1}} will equal \VS{\keyt_{\sss 1}} because the initial FS-PRG's state is the same (from the setup phase) and the two generators have been updated the same number of times. The client then encrypts the verifier and challenge under \VC{\keyt_{\sss 1}} and sends this to the server.
%
%
On receiving and validating this message, the server decrypts the message using \VS{\keyt{\sss 1}}, then extracts the challenge and verifier.
If the challenge does not match the one corresponding to the current protocol exchange, the protocol halts.
If the challenge does match, the server stores the verifier associated with the client's account.

Finally, the device discards the challenge, \VC{\keyt_{\sss 1}}, PIN, and \VC{\verifier} so that the PIN can no longer be recovered from the device. Note that the device can re-generate \VC{\verifier} using \VC{\salt} when the client types in its PIN again. The server also discards the challenge and \VC{\keyt{\sss 1}} as they are no longer needed. Following the successful completion of this protocol, the server will store the verifier corresponding to the client's selected PIN and both server and device will have synchronised their FS-PRG's state.

\subsection{Authentication Phase}
\label{sec:authentication}

The goal of the authentication process is to give the server assurance that the device is currently present, the correct PIN has been entered, and the client has been shown the transaction that the server wishes to execute. 
%

\begin{figure}[!htbp]
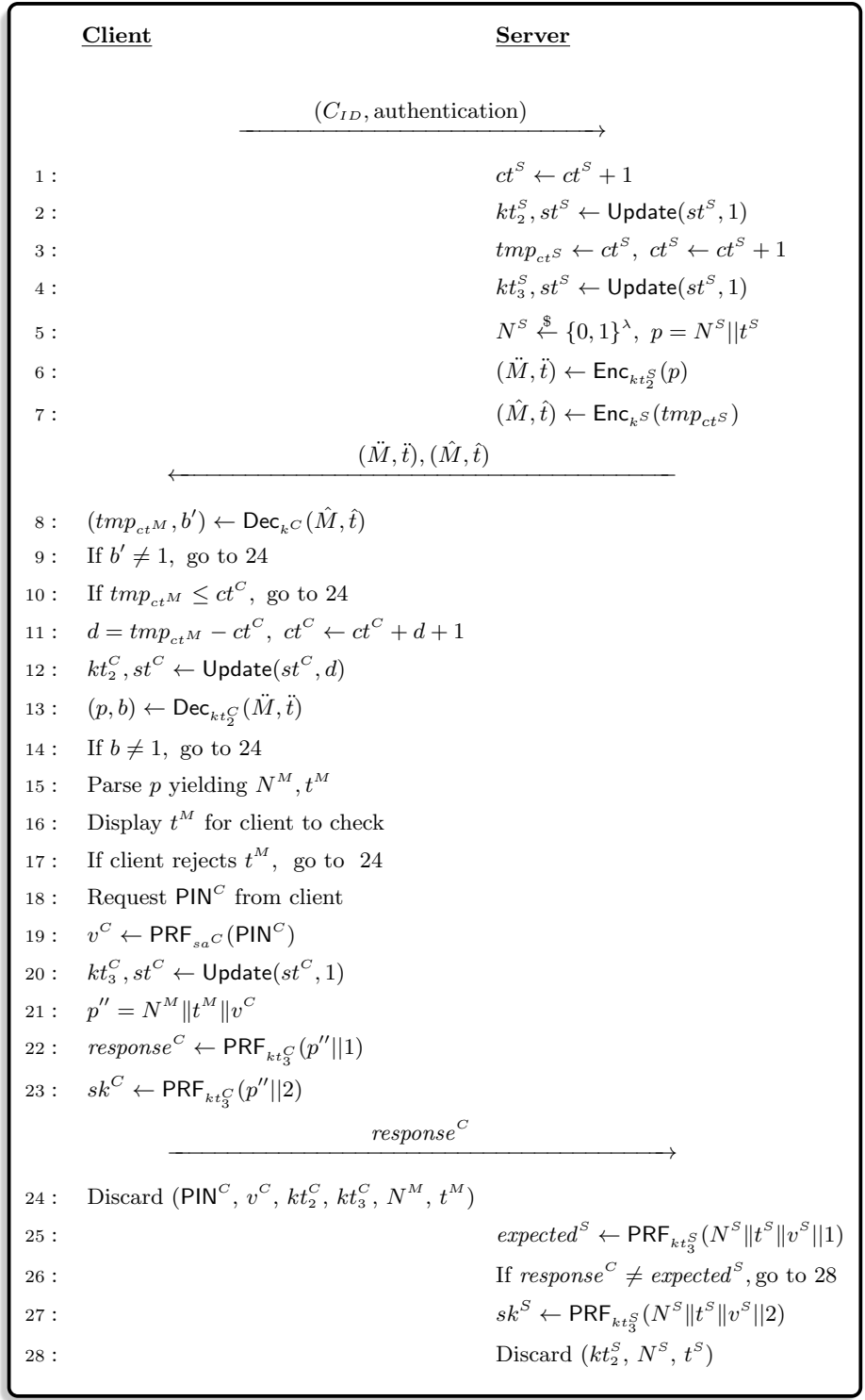

\setlength{\fboxsep}{.8pt}
\begin{center}
     \begin{tcolorbox}[enhanced,width=4.75in, height=197mm, left=1mm,top=0.1mm,
    drop fuzzy shadow southwest,
    colframe=black,colback=white]
 \centering
 \procedure{}{%
  \hspace{8mm}\underline{\textbf{Client}} \< \hspace{0ex} \<\hspace{2.6mm} \underline{\textbf{Server}} \\
\pclb
\sendmessagerightx[5cm]{6}{(C_{\sss ID}, \text{authentication}) } \\
 \pcln \< \<\hspace{2.6mm} \VS{\counter} \gets \VS{\counter} + 1 \\ 
 \pcln \< \<\hspace{2.6mm}  \VS{\keyt_{\sss 2}}, \VS{\state} \gets \update(\VS{\state}, 1)\\
  \pcln \< \<\hspace{2.6mm}  \tmp_{\sss\VS{\counter}} \gets \VS{\counter}, \ \VS{\counter} \gets \VS{\counter} + 1 \< \< \\
 %
 %
 \pcln \< \< \hspace{2.6mm} \VS{\keyt_{\sss 3}}, \VS{\state} \gets \update(\VS{\state}, 1)\\
 \pcln \< \<\hspace{2.6mm} \VS{\nonce} \stackrel{\$}\leftarrow \bin^{\sss \secpar}, \  p= \VS{\nonce} || \VS{\trans} \ \\  
 \pcln \< \<\hspace{2.6mm} (\ddot M, \ddot t)\leftarrow \enc_{\sss\VS{\keyt_{\sss 2}}}(p)\\
 \pcln \< \<\hspace{2.6mm} (\hat M, \hat t)\leftarrow \enc_{\sss\VS{k}}( \tmp_{\VS{\counter}})\pclb
 \sendmessageleftx[7cm]{6}{(\ddot M, \ddot t), (\hat M, \hat t)} \\ 
 %
 %
   \pcln (\tmp_{\sss\VM{\counter}},b') \gets \mathsf{Dec}_{\sss\VC{k}}(\hat M, \hat t) \< \< \\
  \pcln \iif b'\neq 1,    \text{\ go\ to\ } \ref{auth:fail} \\ 
    \pcln \iif \tmp_{\sss\VM{\counter}} \le \VC{\counter}, \text{\ go\ to\ } \ref{auth:fail} \< \< \\
   %
 \pcln d=  \tmp_{\sss\VM{\counter}}-\VC{\counter},\  \VC{\counter} \gets \VC{\counter} + d+1\  \< \< \\
 %
    \pcln \VC{\keyt_{\sss 2}},  \VC{\state} \gets \update(\VC{\state}, d) \< \< \\ 
   %
   %
 %
  \pcln (p,b) \gets \mathsf{Dec}_{\sss\VC{\keyt_{\sss 2}}}(\ddot M, \ddot t) \< \< \\
  \pcln \iif b\neq 1,    \text{\ go\ to\ } \ref{auth:fail} \\ 
  \pcln \text{Parse\ } p \text{\ yielding\ } \VM{\nonce}, \VM{\trans}  \< \< \\ 
 \pcln \text{Display \VM{\trans} for client to check} \< \< \\
 \pcln \text{\iif client rejects \VM{\trans}, \text{\ go\ to\ } \ref{auth:fail}} \< \< \\
 %
 %
 \pcln \text{Request \VC{\pin} from client} \< \< \\
 \pcln \VC{\verifier} \gets \prf_{\sss\VC{\salt}}(\VC{\pin}) \< \< \\
 \pcln  \VC{\keyt_{\sss 3}}, \VC{\state} \gets \update(\VC{\state}, 1) \< \< \\
 \pcln p''=  \VM{\nonce} \|  \VM{\trans} \| \VC{\verifier}\< \< \\ 
 \pcln \label{auth:gen-res}  \VC{\mathit{response}} \gets \prf_{\sss\VC{\keyt_{\sss 3}}}(p''|| 1) \< \< \\ 
  \pcln sk^{C} \gets \prf_{\sss\VC{\keyt_{\sss 3}}}(p''|| 2) \< \<  \pclb
 %
 \sendmessagerightx[7cm]{6}{ \VC{\mathit{response}}} \\
  \pcln \label{auth:fail} \text{Discard (\VC{\pin}, \VC{\verifier}, \VC{\keyt_{\sss 2}}, \VC{\keyt_{\sss 3}}, \VM{\nonce}, \VM{\trans}}) \\ 
 \pcln \< \<\hspace{2.6mm} \VS{\mathit{expected}} \gets \prf_{\sss\VS{\keyt_{\sss3}}}(\VS{\nonce}  \| \VS{\trans} \| \VS{\verifier}||1) \\
 \pcln \< \<\hspace{2.6mm} \iif \VC{\mathit{response}} \ne \VS{\mathit{expected}}, \text{go\ to\ } \ref{auth:fail-server} \\
\pcln\< \<\hspace{2.6mm}  sk^{S} \gets\prf_{\sss\VS{\keyt_{\sss3}}}(\VS{\nonce}  \| \VS{\trans} \| \VS{\verifier}|| 2)\\
 %
 %
  \pcln \label{auth:fail-server}\< \<\hspace{2.6mm}  \text{Discard (\VS{\keyt_{\sss 2}}, \VS{\nonce}, \VS{\trans}}) 
 }
\end{tcolorbox}
\end{center}
    \caption{Authentication phase.}
    \label{fig:auth}
\end{figure}

This phase works as follows. The server first updates the FS-PRG's state and corresponding counter, which results in a new state \VS{\state}, a new random value  \VS{\keyt_{\sss 2}}, and a new temporary counter $\tmp_{\VS{\counter}}$. The server updates the state and the counter one more time which yields a new state \VS{\state}, a new random value \VS{\keyt_{\sss 3}}, and a new counter \VS{\counter}. The server generates a random challenge and two ciphertexts, $\ddot M$ and $\hat M$. The former ciphertext consists of the random challenge and the description of the transaction, encrypted under key \VS{\keyt_{\sss 2}}. The latter ciphertext contains the counter $\tmp_{\VS{\counter}}$, encrypted under key $\VC{k}$. The reason $\tmp_{\VS{\counter}}$ is encrypted under key $\VC{k}$ is to allow the device to decrypt the ciphertext easily 
in case of previous message loss; for instance, when the server sends $(\ddot M, \hat M)$ to the server, but they are lost in transit, multiple times, and a fresh pair finally arrives the client after the server sends them upon the client's request. Encrypting $\tmp_{\VS{\counter}}$ under key $\VC{k}$ (instead of one of the evolving keys) lets the device deal with such a situation.

Upon receiving the ciphertexts, the device validates and decrypts the messages. It extracts the challenge \VM{\nonce}, counter $\tmp_{\VS{\counter}}$, and transaction \VM{\trans}. It ensures that its own counter is behind the received counter. As will be discussed in \prettyref{sec:synchronisation}, this check should always succeed. The device synchronises its state and counter using the server's messages. Next, the device displays the transaction for the client to check. If the client does not accept the transaction (e.g., due to an attempted man-in-the-browser attack), then the protocol aborts immediately. Assuming the client is willing to proceed, then the device prompts for the PIN, and computes the verifier \VC{\verifier} using the key \VC{\salt}. If the client enters the correct PIN, the verifier will be the same as the one sent to the server during the enrolment phase.


For the device to generate the response message, first it updates its state one more time, which results in a pseudorandom value \VC{\keyt_{\sss 3}}. Then, it derives a pseudorandom value, \VC{\mathit{response}}, from a combination of the random challenge \VM{\nonce}, transaction \VM{\trans}, verifier \VC{\verifier}, and $x=1$  using $\prf$ and \VC{\keyt_{\sss 3}}. The device generates a session key, using the above combination and key with a difference that now $x=2$. The response message is sent to the server. The device discards the PIN, the verifier, all FS-PRG keys, the challenge, and the transaction's description, so as to protect the PIN from discovery. The server computes the expected response message based on its own values of the challenge, transaction, and verifier. Note that the verifier is retrieved from the value set during the enrolment phase. The server then compares the expected response with the response sent by the client. Only if these match, the authentication is considered to have succeeded. If the response does not match the one the server expects this could indicate that the message was tampered with, or that the client entered an incorrect PIN. Next, the server generates the session key the same way as the device does.  The server also discards the FS-PRG key, the challenge, and the transaction's description.

Below, we formally state the security of our protocol.  First, we present a theorem stating that the advantage of an adversary in breaking the semantic security of the above protocol is negligible.  
\begin{theorem}[Semantic Security]
Let $\adv$ be a probabilistic polynomial time (PPT) adversary with less than $q_{\sss s}$ interactions with the parties and $q_{\sss p}$ passive eavesdropping, i.e., number of local executions. Let $\lambda$ be a security parameter and $Adv_{\sss\pi}^{\sss ss}(\A)$ be  $\adv$'s advantage (in breaking the semantic security of an AKE scheme $\pi$) as defined in Section \ref{sec::model}. Then, such an advantage for the protocol $\psi$ has the following upper bound:  
\begin{equation*} 
Adv_{\sss \psi}^{\sss ss}(\A) \leq 2(q_{\sss s}+q_{\sss p})\Big(Adv^{\sss\prf}(\adv)+Adv^{\sss Enc}(\adv)\Big)+\frac{8(2q_{\sss s}+q_{\sss p})}{2^{\sss\lambda}}
\end{equation*}
\end{theorem}

Next, we present a theorem stating that the advantage of an adversary in breaking the authentication of the above protocol is negligible.  

\begin{theorem} [Authentication]
Let PIN be an element distributed uniformly at random over a finite dictionary of size $N$. Also, 
let $\adv$ be a PPT adversary with less than $q_s$ interactions with the parties and $q_p$ passive eavesdroppings. Let $\lambda$ be a security parameter and $Adv_{\sss\pi}^{\sss aut}(\A)$ be  $\adv$'s advantage (in breaking the authentication of an AKE scheme $\pi$) as defined in Section \ref{sec::model}. Then, in the protocol $\psi$, $Adv_{\sss \psi}^{\sss aut}(\A)$ has the following upper bound:  
  \begin{equation*}
 Adv_{\sss \psi}^{\sss aut}(\A)  \leq (q_{\sss s} + q_{\sss p})\Big(Adv^{\sss\prf}(\adv)+Adv^{\sss Enc}(\adv)\Big)+\frac{9q_{\sss s}+4q_{\sss p}}{2^{\sss\lambda}}+  \cfrac{q_{\sss s}}{N}
 \end{equation*}
\end{theorem}

\section{Informal Security Analysis}
\label{sec:security}

In this section, we informally analyse the security of the proposed protocol. We analyse its security through five scenarios defined in terms of adversary capabilities and protection goals.
The scenarios are designed to assume a strong adversary so that the results are generalisable to other situations, but are constrained so as to make sense, e.g., we assume that at least one factor is secure.


\subsection{Threats and Protection objectives}
In this section,  we first list a set of threats that our protocol must resist. 



\begin{itemize}
\item[$\bullet$]\textit{T.DEV: Device access.} An adversary may steal the authentication device. The adversary will then know \VC{k}, \VC{\salt} and the current values of \VC{\counter} and \VC{\state}, because we assume the device does not take advantage of a trusted chipset. The adversary does not learn \VC{\verifier}, \VM{\trans}, \VC{\pin}, \VC{\keyt_{\sss 1}}, \VC{\keyt_{\sss 2}}, \VC{\keyt_{\sss 3}} or previous values of \VC{\state}; as these are all discarded at the end of a protocol exchange. We assume that the client will not use the device after it has been stolen, and will be issued with a replacement.

\item[$\bullet$]\textit{T.MITM: Man-in-the-middle.} An adversary may have access to the traffic exchanged between the client and server.

\item[$\bullet$]\textit{T.PIN: Knowledge of PIN.} An adversary may know the PIN entered by a client, for example from observing them type it in.

\item[$\bullet$]\textit{T.SRV: Server compromise.} The server is the party relying on the authentication, so it does not make sense for the server to be wholly malicious.
However, it is reasonable to believe that the server database could be compromised, disclosing \VS{k}, \VS{\verifier}, and the current values of \VC{\counter} and \VC{\state}.

\end{itemize}


Next, we present the high-level security objective that our protocol must achieve. 

\begin{itemize}
\item[$\bullet$]\textit{O.AUTH: Authentication.} If the server considers the authentication to have succeeded then the correct device was used and the correct PIN was entered.

\item[$\bullet$]\textit{O.TRAN: Transaction authentication.} If the server considers the authentication to have succeeded then the correct device was used, the correct PIN was entered, and the device showed the correct transaction.

\item[$\bullet$]\textit{O.PIN: PIN protection.} The adversary should not be able to discover the client's PIN. 
\end{itemize}

\subsection{Scenarios}
In this section, we briefly explain why the protocol meets its objective in different threat scenarios. 

\begin{enumerate}

\item\textit{O.AUTH against 
T.PIN and T.MITM.}  The first scenario we consider is the case where the adversary does not have
access to the authentication device but does know the client's PIN and communication between the client and server.  For the adversary to perform a successful authentication, it must compute $\prf_{\sss\VC{\keyt_{\sss 3}}}(\VM{\nonce} \|  \VM{\trans} \| \VC{\verifier}|| 1)$. However, it does not know \VC{\keyt_{\sss 3}} or the state from which \VC{\keyt_{\sss 3}} has been generated; since  \VC{\keyt_{\sss 3}} is an output of $\prf$ and is sufficiently large, it is computationally indistinguishable from a truly random value.  The probability of finding it is negligible in the security parameter.  Thus, the only party which will generate a valid response is the device itself (when the PIN is provided) at line \ref{auth:gen-res} of \prettyref{fig:auth}.  We have already assumed that the adversary does not have access to the device; therefore, it cannot generate a valid response.

\item\textit{O.AUTH against T.DEV and T.MITM.}\label{O.AUTH:T.DEV-plus-T.MITM}
In this scenario, the adversary has compromised the client's device (but not its PIN), has records of previous messages, and wishes to impersonate the client.
%
Since the random challenge in the expected response is unique, and the $\prf$ provides an unpredictable output, previous responses will not be valid; so, a reply attack would not work. The adversary can use the device to discover ($\VC{\keyt_{\sss 2}}, \VC{\keyt_{\sss 3}}$) and all the parameters of the response message, except the PIN. In this case, it has to perform an online dictionary attack by guessing a PIN, using the extracted parameters to generate a response, and sending the response to the server. But, the server will lock out the device if the number of incorrect guesses exceeds the predefined threshold. Other places where the PIN is used are in (i) the enrolment response, where the verifier derived from the PIN is encrypted under an evolving fresh secret key, and (ii) the authentication response, where the response is a pseudorandom value derived from the PIN's verifier using an evolving fresh secret key. In both cases, the evolving keys cannot be obtained from the current state, due to the security of FS-PRG.

\item\textit{O.PIN against T.DEV and  T.MITM.}
The adversary has compromised the device and wishes to obtain the client's PIN. As with Scenario \ref{O.AUTH:T.DEV-plus-T.MITM}, the PIN cannot be obtained from the device, the responses in the authentication or enrolment phases.

\item\textit{O.PIN against T.SRV and T.MITM.}
\label{sec:servercompromise}
The adversary has compromised the server and wishes to obtain the client's PIN.
In this case, the adversary has learned the verifier but does not know the value of the secret key, used to generate the verifier. If the server retains values of the verifier for previous PINs (in the case where the server does not delete them), then the adversary would also learn further verifiers for the same device. The PIN is only used for computing the verifier, so the only way to obtain the PIN would be to find the key of the $\prf$ which is not possible except for a negligible probability in the security parameter. The only information this discloses is that if two values for the verifier are equal, then that implies that two PINs for the same device were equal. Even this minimal information leakage can be removed if the server rejects the PINs that were used before. 

\item\textit{O.TRAN against T.PIN and T.MITM or T.DEV and T.MITM.} As we discussed above, an adversary cannot successfully authenticate, even if it sees the traffic between the client and server and has access to either the PIN or the device. Furthermore, due to the security of the authenticated encryption, the device can detect (except for a negligible probability) if the transaction's description, that the server sends to it, has been tampered with. 
 
\end{enumerate}

\subsection{Excluded Scenarios}

We exclude some scenarios that do not make sense or are not possible to secure against.

\begin{itemize}
\item[$\bullet$]\textit{Compromised PIN and device.}
If the adversary has compromised both factors of a 2FA solution, then the server cannot distinguish between the adversary and the legitimate client.

\item[$\bullet$]\textit{Authentication on server compromise.}
If the adversary has compromised the server, then it can either directly perform actions of the server or change keys to ones known by the adversary. Therefore, it does not make sense to aim for O.AUTH in this situation.

\item[$\bullet$]\textit{Compromised server and device.}
If the adversary has compromised the server and the device, then the PIN can be trivially brute-forced with knowledge of \VS{\verifier} and \VC{\salt}. 
\end{itemize}

\section{Synchronisation}
\label{sec:synchronisation}

A user's device needs to be synchronised with the server
in order for the server to check the correctness of the response generated by the device.
This is particularly the case in our proposed protocol because
if one side advances too far, it is by design impossible for it to
move backwards. Specifically, we must provide assurance that the server state remains at the same state as the client's state, or that the server is ahead of the client, i.e., $\VS{\counter} \geq \VC{\counter}$. Then, as challenge messages always contain the current value of the server's counter, the client is always able to catch up with the server. We achieve this via three approaches.  Firstly, by requiring the FS-PRG's state to advance with the counter, such that the counter is consistent with the state. Secondly, by requiring that the client never advances its state directly, but only advances to the point that the server currently is at. Thirdly, by requiring the client only to advance its state in response to an authenticated challenge from the server.

The protocol takes into account the case where messages are dropped.
Response messages are not involved in advancing the forward-secure state; therefore, if these messages are dropped, then it would not have any effect on synchronisation. However, challenge messages are important, if any of them is dropped, then the device would not advance the state and would be behind the server. Nevertheless, this would not cause any issue, because the server's next challenge message will include the new value of the counter and the device will advance the state until it matches the server's state. Note that 
the FS-PRG advance process is fast; thus,  multiple invocations of this will not create a noticeable delay.
 Note that in the case where the enrolment's response message is dropped, the PIN will remain unchanged; as a result,  the client may be surprised that the new PIN does not work. But, the old PIN will keep working and enrolment can be repeated to update the PIN.


\section{System Usability}
\label{sec:asymmetric}

%

\begin{figure}
    \centering
    \includegraphics[width=6cm]{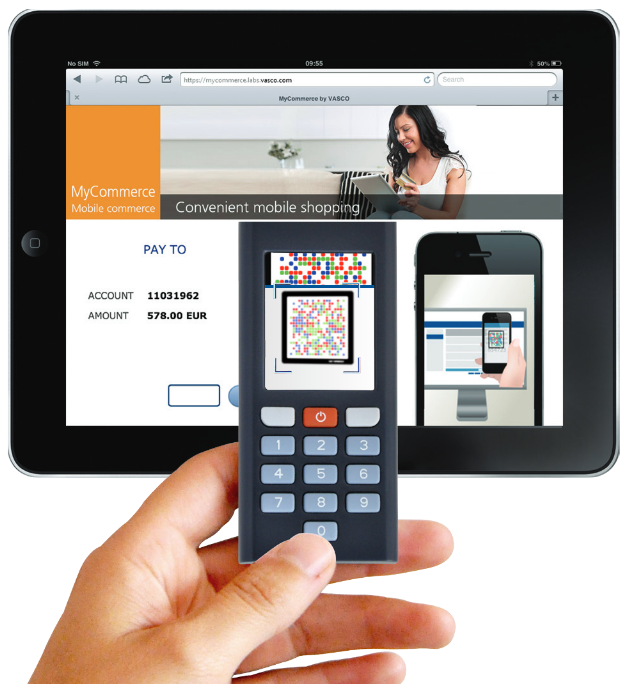}
    \caption{OneSpan Digipass 770 authentication device.}
    \label{fig:dp770}
\end{figure}

Usability is of critical importance for an effective authentication system as otherwise, clients will refuse to use it or implement insecure workarounds \cite{de2013comparative}.
As we highlighted in Section \ref{sec::model}, in our protocol, the server interacts with the device via the client. To accommodate usability and let the device easily receive the server's message, we require the device to be equipped with a scanner/camera, such as the one shown in \prettyref{fig:dp770}.

In our protocol, the encrypted messages the server sends are in a form of a QR code, so the client can easily scan them from its computer's screen and transfer them to the device which verifies and decrypts the input messages, as explained in the protocol. However, to avoid limiting the protocol's application, we do not require the client's computer to be equipped with a scanner/camera. Therefore, the client needs to manually insert the message output of the device into its computer.  Once again, to improve usability, it is desirable to reduce the size of each message that the device outputs. {Our protocol allows truncating the verifier, as it is resistant to offline brute-force attacks, with a caveat. Truncating the verifier will create collisions such that for some values of the verifier there will be multiple PINs which are valid. Consequently, an adversary who has stolen a device has a better chance of guessing the PIN than the ideal case where the original message is used and its chance is negligible in the security parameter or the PIN's universe size.} This security-usability trade-off is not specific to our protocol and exists in all hardware token-based multi-factor authentication protocols that do not assume clients' computers are equipped with a scanner and clients have to insert the device's messages into their computer.

Another consideration is handling mistyped or forgotten PINs. As we highlighted in Section \ref{sec::the-protocol}, the device can ask the client to confirm its PIN by entering the PIN twice; then the device compares the two entries and alters the client if they do not match.  However, it does not check and alert the client whether the PIN is correct.  Such a check has to be carried out on the server-side which comes at a cost of decreased usability.  The protocol presented in this work is useful in its own right; however, it could be considered a basis for a more user-friendly system, tailored to a particular scenario in which it is to be used.



%


\section{Evaluation}\label{sec::eval}

In this section, we analyse and compare the 2FA protocol, we presented in Section \ref{sec::the-protocol}, with the smart-card-based protocol proposed in  \cite{WangW18} and the hardware token-based protocol in \cite{JareckiJKSS21} because the latter two protocols are relatively efficient, do not rely on secure chipsets, and they consider the same security threats as we do, e.g., resistance against card/token loss, against an offline attack, against a  corrupt server. We summarise the result of the analysis in Table \ref{comparisonTable}. 



\begin{table} 

\begin{center}
\caption{ \small Comparison of efficient two-factor authentication protocols.}  \label{comparisonTable} 
\begin{tabular}{|c|c|c|c|c|c|c|c|c|c|} 
\hline

{\cellcolor{gray!40}\scriptsize {Features}} &\cellcolor{gray!40}{\scriptsize {Operation}}&\cellcolor{gray!40}{ \scriptsize {Our Protocol}}&\cellcolor{gray!40}{\scriptsize{\cite{WangW18}}}&\cellcolor{gray!40} {\scriptsize{\cite{JareckiJKSS21}}}  \\
\hline

\multirow{2}{*}{\scriptsize Computation cost}

&\scriptsize{Sym-key}&\cellcolor{gray!20}\scriptsize$18$&\cellcolor{gray!20}\scriptsize$19$&\cellcolor{gray!20}\scriptsize $7$\\

\cline{2-5}

&\scriptsize {Modular expo.}&\cellcolor{gray!20}\scriptsize {$0$}&\cellcolor{gray!20}\scriptsize$5$&\cellcolor{gray!20}\scriptsize$12$\\

\hline

\scriptsize  Communication cost&$-$&\cellcolor{gray!20}\scriptsize$2804$-bit&\cellcolor{gray!20}\scriptsize$3136$-bit&\cellcolor{gray!20}\scriptsize$3900$-bit\\

\hline 
\scriptsize Not requiring multiple pass/PIN&$-$&\cellcolor{gray!20}\scriptsize{\textcolor{blue}\checkmark}&\cellcolor{gray!20}\scriptsize\textcolor{red}{$\times$}&\cellcolor{gray!20}\scriptsize\textcolor{red}{$\times$}\\ 
\hline
\scriptsize Not requiring modular expo.  &$-$&\cellcolor{gray!20}\scriptsize{\textcolor{blue}\checkmark}&\cellcolor{gray!20}\scriptsize\textcolor{red}{$\times$}&\cellcolor{gray!20}\scriptsize\textcolor{red}{$\times$}\\ 

\hline

\scriptsize Security assumption  &$-$&\cellcolor{gray!20}\scriptsize{Standard}&\cellcolor{gray!20}\scriptsize{Random oracle}&\cellcolor{gray!20}\scriptsize{Random oracle}\\ 

\hline
\end{tabular}
\end{center}
\end{table}


\subsection{Computation Cost}

We start by analysing our protocol's computation cost. First, we focus on the protocol's enrolment phase. The client's computation cost, in this phase, is as follows. It invokes the authenticated encryption scheme $2$ times. It also invokes once the pseudorandom function, $\prf$.
%
%
  Moreover, the server invokes the authenticated encryption scheme twice, and calls $\prf$ only once, in this phase. 
%
%
Now, we move on to the authentication phase. The client invokes the authenticated encryption scheme $2$ times and invokes $\prf$ $4$ times. 
%
%
 In this phase, the server invokes the authenticated encryption scheme and $\prf$ $2$ and $4$ times respectively. 

Next, we analyse the computation cost of the protocol in \cite{WangW18}. We consider all operations performed on the smart card or card reader as client-side operations. The enrolment phase involves $3$ and $2$ invocations of a hash function at the client and server sides respectively. This protocol has an additional phase called login which costs the client  $5$ invocations of the hash function and $2$ modular exponentiations for each authentication.  The verification requires the server $6$ invocations of the hash function and $2$ modular exponentiations. This phase requires the client to perform $1$ modular exponentiation and invoke the hash function $3$ times. 


Now, we analyse the computation cost of the protocol presented in \cite{JareckiJKSS21}. In our analysis, due to the high complexity of this protocol, we estimate the protocol's \emph{minimum} costs. The actual cost of this protocol is likely to be higher than our estimation. The protocol's phases have been divided into enrolment and login, i.e., verification. The enrolment phase requires a client to perform single modular exponentiation and invoke a hash function $2$ times. It also involves, as a subroutine, the initialisation of asymmetric  ``password-authenticated key exchange'' (PAKE) proposed in \cite{GentryMR06}, which involves at least $2$ modular exponentiations, $1$ invocation of hash function and symmetric key encryption. In the login phase, the client performs at least $7$ modular exponentiations. In the login phase, the server invokes a pseudorandom function once and performs at least $2$ modular exponentiations and $2$ symmetric-key encryptions (due to the execution of PAKE).

Thus,  our protocol and the ones in \cite{WangW18,JareckiJKSS21} involve a constant number of symmetric key primitive invocations; however, our protocol does not involve any modular exponentiations, whereas the protocol in  \cite{WangW18,JareckiJKSS21} involves a constant number of modular exponentiations which leads to a higher cost. 

\subsection{Communication Cost}

We first analyse our protocol's communication cost. In the enrolment phase,  the client only sends two pairs of messages: $(C_{\text ID}, $ $\text{enrolment})$ and $(M', t')$, where the total size of messages in the first pair is about $250$ bits (assuming the ID is of length $128$ bits),  while the total size of messages in the second pair is about  $512$ bits as they are the outputs of symmetric-key primitives, i.e., symmetric key encryption and message authentication code schemes whose output size is $256$ bits. The server sends out only a single pair $(M', t')$ whose total size is about $512$ bits. 
The parties' communication cost in the authentication phase is as follows. The client only sends three messages: $(C_{\text ID}, $ $\text{authentication}, \VC{\mathit{response}})$, where the combined size of the first two messages is about $250$ bits while the third message's size is about $256$ bits. The server sends only two pairs of messages $(\ddot M, \ddot t)$ and $(\hat M, \hat t)$ with a total size of $1024$ bits. Therefore, the total communication cost that our protocol imposes is about $2804$ bits. 

Next, we evaluate the cost of the protocol in \cite{WangW18}. The client's total communication cost in the enrolment and login phases is $1792$ bits. Note that we set the client's ID's size to $128$ bits and we set the hash function output size to $160$ bits, as done in \cite{WangW18}. In the verification phase, the client sends to the server a single value of size $160$ bits. In the verification, the server sends to the client two values that in total costs the server $1184$ bits. So, this protocol's total communication concrete cost is about $3136$ bits.

Now, we analyse the communication cost of the protocol in \cite{JareckiJKSS21}. As before, in our cost evaluation, we estimate the protocol's minimum cost.  In the enrolment phase, a client sends a random key, of a pseudorandom function, to the server and the device, where the size of the key is about $128$ bits. It also, due to the initialisation of PAKE, sends a $128$-bit value to the server. In the login phase, the client sends out three parameters of size $128$ bits and a single parameter of size $20$ bits.  It also invokes PAKE with the server that requires the client to send out at least one signature of size $1024$ bits. The device sends to the client a ciphertext of asymmetric key encryption which is of size $1024$ bits along with a $20$-bit message. Thus, the client-side total communication cost is at least $2856$ bits. The server in the login phase sends out a message $zid$ of size $20$ bits and invokes PAKE that requires the server to send out at least a ciphertext of symmetric key encryption which is of size $1024$ bits.  So, the server-side communication cost is at least $1044$ bits. So, the total communication cost of this protocol is at least $3900$ bits.

Hence, our protocol imposes a $10\%$ and $40\%$ lower communication cost than the protocols in \cite{WangW18} and \cite{JareckiJKSS21} do respectively.


\subsection{Other Features}

 In our protocol, a client needs to know only a single secret (i.e., a  PIN). Nevertheless, in the protocol in \cite{WangW18} a client requires to know (and insert into the verification algorithm) an additional secret; namely, a secret random ID. Thus, the client needs to remember two secrets in total.  As shown in \cite{Scott12a}, this scheme will not remain secure, even if only the client's ID is revealed. Furthermore, the protocol in \cite{JareckiJKSS21} requires the client to remember or locally store at least one cryptographic secret key of sufficient length, e.g., $128$ bits; this secret key is generated via invocation of a subroutine protocol (called PAKE) and must not be kept on the device. 
 
 Furthermore, our protocol is secure in the standard model while the protocols in \cite{WangW18,JareckiJKSS21} are in the non-standard random oracle model.


\section{Formal Security Analysis}

In this section, we present the security proof of the protocol, presented in Section \ref{sec::the-protocol}.  First, we prove the semantic security of the scheme and then prove its authentication. 

\subsection{Semantic Security}\label{sec::semSec-proof}

In this section, we assert that under standard assumptions protocol $\psi$, presented in Section \ref{sec::the-protocol}, securely distributes
session keys. To do so, we incrementally define a sequence of games starting at the real
game $G_{\sss 0}$ and ending up at $G_{\sss  7}$. We first define various events in every game and then explain each game. 

\begin{itemize}
\item  $S_{\sss  i}$: it takes place if $b=b'$, where $b$ is the bit involved in the test query and $b'$ is the output of $\A$ which wants to guess $b$. 
\item $Auth_{\sss  i}$: it occurs if $\A$ generates and sends to the server an authenticator message that is accepted by the server.
\item $Enc_{\sss  i}$: occurs if $\A$ submits data it has encrypted by itself using the correct key that an honest party would use to encrypt. 
\end{itemize}

\begin{itemize}
\item[$\bullet$] \textit{\textbf{Game}} $G_{\sss  0}$: This is the real attack game. Several oracles are  available to the adversary; namely, the pseudorandom function ($\prf$),  the encryption/decryption oracles ($\mathtt{Enc}$ and $\mathtt{Dec}$) and all instances $C^{\sss  i}$ and $S^{\sss  j}$.
According to the definition we presented in Section \ref{sec::model}, the advantage of the adversary in this protocol is: 
\begin{equation}\label{eq::adv-ake-}
Adv_{\sss  \psi}^{\sss  ss}(\A)=2Pr[S_{\sss  0}]-1
\end{equation}

Similar to the security proof in \cite{BressonCP03}, we assume that if any of the games halts and $\A$ does not output $b'$, then $b'$  is chosen at random. Also, if $\A$ has not finished playing the game after sending $q_{\sss  s}$  \send$(.)$ queries or if it plays the game more than a predefined time $t$, the game is stopped and a random value is assigned to $b'$. 

\item[$\bullet$] \textit{\textbf{Game}}  $G_{\sss  1}$: This game is similar to  $G_{\sss  0}$, except that the output of the $\prf$ is replaced by an output of  a uniformly random function $f$, i.e., when the simulator in Figure \ref{fig::PRF-SIM} is used. Since the output of $f$ (in the simulator) and $\prf$ are indistinguishable, except with a negligible probability, we will have: 
\begin{equation}\label{eq::game_1}
|Pr[S_{\sss  1}]-Pr[S_{\sss  0}]|\leq (q_{\sss s}+q_{\sss p})Adv^{\sss\prf}(\adv)
\end{equation}

that captures both send and execute queries. We highlight that as we use a standard $\prf$, the probability of finding a collision is $0$.

\item[$\bullet$] \textit{\textbf{Game}}  $G_{\sss  2}$: This game is the same as $G_{\sss  1}$, with the difference that we simulate the authenticated encryption scheme (i.e., $Enc$ and $Dec$ algorithms). We replace the output of $Enc$ with a uniformly random value picked from the encryption scheme's range. The adversary has access to the encryption and decryption oracles. Since we treat the encryption scheme as a black box, the two games are distinguishable except with a negligible probability; this we will have: 
\begin{equation}\label{eq::game_2}
|Pr[S_{\sss  2}]-Pr[S_{\sss  1}]| \leq (q_{\sss s}+q_{\sss p}) Adv^{\sss Enc}(\adv)
\end{equation}

The above also captures both the send and execute queries. Since we have used a standard encryption scheme, the probability of finding a collision (e.g., two ciphertexts result in the same plaintext or two plaintexts result in the same ciphertext) is $0$, as the scheme is bijective.

\item[$\bullet$] \textit{\textbf{Game}}  $G_{\sss  3}$: This game is the same as $G_{\sss  2}$, with the difference that we simulate the verification of a transaction, i.e., via predicate $\phi$ defined in Section \ref{sec::model}. Moreover, we simulate all parties' instances via defining simulators for  \send, \execute, \reveal, and \test\ queries. We present the simulators for client's and server's \send\ queries in Figures \ref{fig::Send-sim-to-client} and \ref{fig::Send-sim-to-server} respectively. Also, we present the simulators for the rest of the queries in Figure \ref{fig::sim-for-exe-rev-test}. By definition,  $\phi$ is a deterministic function, given the transaction $t^{\sss  C}$ and policy $\pi$, it always returns the same output as the client does when verifying $t^{\sss  C}$ in the previous game.  Therefore, both $\phi$ and the client would output identical values, given pair $(t^{\sss  C}, \pi)$, meaning that their outputs are indistinguishable in both games.  Given the above argument, we conclude that:  
\begin{equation}\label{eq::game_3}
Pr[S_{\sss  3}]-Pr[S_{\sss  2}]=0
\end{equation}
\item[$\bullet$] \textit{\textbf{Game}}  $G_{\sss  4}$: This game is the same as $G_{\sss  3}$, with the difference that when the adversary manages to use the correct encryption key and encrypts (or decrypts) a message itself, then the simulation aborts. Therefore, we have: 
\begin{equation*}
|Pr[S_{\sss  4}]-Pr[S_{\sss  3}]|\leq Pr[Enc_{\sss  4}]
\end{equation*}

We know that the key has been picked uniformly at random and is of length $\lambda$ bits (recall that the outputs of $\prf$ have been replaced with truly random values in $G_{\sss  1}$). Therefore:
 $$Pr[Enc_{\sss  4}]=\frac{4(q_{\sss  s}+q_{\sss  p})}{2^{\sss \lambda}}$$ and 
 \begin{equation}\label{eq::game_4}
 |Pr[S_{\sss  4}]-Pr[S_{\sss  3}]|\leq \frac{4(q_{\sss  s}+q_{\sss  p})}{2^{\sss \lambda}}
 \end{equation}

\item[$\bullet$] \textit{\textbf{Game}}  $G_{\sss  5}$: In this game, we modify the simulator such that it would abort if
the adversary correctly guesses the authenticator. Therefore, we modify the way the server responds to  query {\send($S^{\sss  j},   \bar{\VC{\mathit{response}}}$)} as follows: 

\begin{enumerate}
\item computes $\VS{\mathit{expected}} \gets \prf_{\sss\bar{\VS{\keyt_{\sss3}}}}(\ddot{\VS{\nonce}}  \| \VS{\trans} \| \VS{\verifier}|| 1)$.
\item checks if $\bar{\VC{\mathit{response}}}=\VS{\mathit{expected}}$. It proceeds to the next step if the equation holds. 
\item\label{Game::check-sim} checks if $\Big((C_{\sss  ID},  \text{enrolment}), (C_{\sss  ID},  \text{authentication}), (\bar M, \bar t),$ $ (\bar M',$ $ \bar t'),$ $ (\ddot M', $ $\ddot t'), $ $(\hat M', \hat t'), \bar{\VC{\mathit{response}}}\Big)\in \vv L$.

\item\label{Game::prf} checks if $ \bar{\VC{\mathit{response}}}\in L_{\sss \adv}$. 
\item if both checks in steps \ref{Game::check-sim} and \ref{Game::prf} fail, then it rejects authenticator $\bar{\VC{\mathit{response}}}$ and terminates without accepting the key. Otherwise, it accepts the key. 
\end{enumerate}
This game ensures that if the message (i.e., the authenticator) does not come from the simulator or the adversary (which decrypted $\ddot M',\ddot t', \hat M'$, and  $\hat t'$, then correctly computed a valid authenticator by querying $\update$ and $\prf_{\sss  k'}$) then it aborts. So, games $G_{\sss  4}$ and $G_{\sss  5}$ are indistinguishable unless the server rejects a valid authenticator. However, this means the adversary has correctly guessed the output of $\prf$. Thus, 
\begin{equation}\label{eq::game_5}
|Pr[S_{\sss  5}]-Pr[S_{\sss  4}]|\leq \frac{q_{\sss  s}}{2^{\sss \lambda}}
\end{equation}

\item[$\bullet$] \textit{\textbf{Game}}  $G_{\sss  6}$: 
 In this game, we modify the simulator in a way that it would abort if $\adv$ decrypts $ (\ddot M', \ddot t', \hat M', \hat t')$ and uses the result to generate and send a valid authenticator to the server. To do so, we modify the way the server responds to  query {\send($S^{\sss  j},   \bar{\VC{\mathit{response}}}$)}, as follows: 

\begin{enumerate}
\item computes $\VS{\mathit{expected}} \gets \prf_{\sss\bar{\VS{\keyt_{\sss3}}}}(\ddot{\VS{\nonce}}  \| \VS{\trans} \| \VS{\verifier}||1)$.
\item checks if $\bar{\VC{\mathit{response}}}=\VS{\mathit{expected}}$. It proceeds to the next step if the equation holds. 
\item\label{Game::check-sim-} checks if $\Big((C_{\sss  ID},  \text{enrolment}), (C_{\sss  ID},  \text{authentication}), (\bar M, \bar t),$ $ (\bar M',$ $ \bar t'),$ $ (\ddot M', $ $\ddot t'), $ $(\hat M', \hat t'), \bar{\VC{\mathit{response}}}\Big)\in \vv L$. If this check fails, then it rejects authenticator  $\bar{\VC{\mathit{response}}}$ and terminates, without accepting any key.

\item\label{Game::prf-} checks if $(\ddot{\VS{\nonce}}  \| \VS{\trans} \| *,\bar{\VC{\mathit{response}}})\in L_{\sss \adv}$.  
\item aborts, if the above check (in step \ref{Game::prf-}) passes.
\end{enumerate}

The above modification ensures that all valid authenticators are sent by the simulator. Let $\hat {Auth}_{\sss  6}$ be the event that the check in step \ref{Game::prf-} passes. Games $G_{\sss  5}$ and $G_{\sss  6}$ are indistinguishable unless $\hat {Auth}_{\sss  6}$ occurs. Hence,
$$|Pr[S_{\sss  6}]-Pr[S_{\sss  5}]|\leq Pr[\hat {Auth}_{\sss  6}]$$

We know that $\hat {Auth}_{\sss  6}$ occurs with probability $ \frac{q_{\sss  s}}{2^{\sss \lambda}}$ when the query $q=\ddot{\VS{\nonce}}  \| \VS{\trans} \| *$ to $\prf$ results in $\bar{\VC{\mathit{response}}}$. Thus, 
\begin{equation}\label{eq::game_6}
|Pr[S_{\sss  6}]-Pr[S_{\sss  5}]|\leq \frac{q_{\sss  s}}{2^{\sss \lambda}}
\end{equation}

\item[$\bullet$] \textit{\textbf{Game}}  $G_{\sss  7}$: In this game, we modify the simulator such that it would abort if the adversary comes up with the authenticator and session key without decrypting  $ (\ddot M', \ddot t', \hat M', \hat t')$.  Therefore, we modify  the way the client processes query \send($C^{\sss  i}, (\ddot M', \ddot t'), (\hat M', \hat t')$) as follows.
 \begin{enumerate}
 \item compute $\bar{\VC{\mathit{response}}} \gets \prf_{\sss  \ddot t'}( \ddot M'|| 1)$.
 \item compute $\bar {sk}^{\sss C}\gets \prf_{\sss  \ddot t'}( \ddot M'|| 2)$.
 \end{enumerate}

We also amend the way the server compiles query \send($S^{\sss  j},   \bar{\VC{\mathit{response}}}$) as follows. 
\begin{enumerate}[label=\alph*]
\item\label{Game::prf--}  checks if  $(\ddot M'||1, \bar{\VC{\mathit{response}}})\in L_{\sss \adv}$ or  $(\ddot M'||2, \bar {sk}^{\sss C})\in L_{\sss \adv}$. 
\item aborts, if either of the above checks (in step \ref{Game::prf--}) passes.
\end{enumerate}

 Let $\hat {Auth}_{\sss  7}$ be the event that the check in step \ref{Game::prf--} passes. Games $G_{\sss  6}$ and $G_{\sss 7}$ are indistinguishable unless $\hat {Auth}_{\sss  7}$ occurs. Therefore,
$$|Pr[S_{\sss  7}]-Pr[S_{\sss  6}]|\leq Pr[\hat {Auth}_{\sss  7}]$$

Event $\hat {Auth}_{\sss  7}$ occurs with probability $ \frac{2q_{\sss  s}}{2^{\sss \lambda}}$ when the query $q=\ddot M'||1$ to $\prf$ results in $\bar{\VC{\mathit{response}}}$ or $q=\ddot M'||2$ to $\prf$ results in $\bar {sk}^{\sss C}$. Thus, 
\begin{equation}\label{eq::game_7}
|Pr[S_{\sss  7}]-Pr[S_{\sss  6}]|\leq \frac{2q_{\sss  s}}{2^{\sss \lambda}}
\end{equation}

Moreover, the session key and authenticator are random values, as they are the outputs of $\prf$ whose secrete key is not known. Therefore, $Pr[S_{\sss  7}]=\frac{1}{2}$.

By summing up all the above relations \ref{eq::game_1}-\ref{eq::game_7}, we would have 
\begin{equation}\label{equ::sum}
|Pr[S_{\sss  7}]-Pr[S_{\sss  0}]|\leq (q_{\sss s}+q_{\sss p})\Big(Adv^{\sss\prf}(\adv)+Adv^{\sss Enc}(\adv)\Big)+\frac{4(q_{\sss  s}+q_{\sss  p})}{2^{\sss \lambda}}+\frac{4q_{\sss  s}}{2^{\sss \lambda}}
\end{equation}

By combining Equations \ref{eq::adv-ake-} and \ref{equ::sum}, we would have: 
\begin{equation*} 
Adv_{\sss  \psi}^{\sss  ss}(\A) \leq 2(q_{\sss s}+q_{\sss p})\Big(Adv^{\sss\prf}(\adv)+Adv^{\sss Enc}(\adv)\Big)+\frac{8(2q_{\sss  s}+q_{\sss  p})}{2^{\sss \lambda}}
\end{equation*}

This completes the proof.

%


\end{itemize}


\begin{figure}[!hpt]
\begin{center}
    \begin{tcolorbox}[enhanced,width=4.7in, left=.1cm,
    drop fuzzy shadow southwest,
    colframe=black,colback=white]
{\small{
\doubleunderline{Pseudorandom function}\\

The simulator upon receiving query $(\prf,q)$ acts as follows. 

\begin{itemize}
 \item picks a function $f$, i.e., $f\stackrel{\sss \$}\leftarrow Func$, where $Func$ is the set of all functions mapping $|q|$-bit strings to $|q|$-bit strings. 
 \item adds record $(q,f(r))$ to list $L_{\sss\adv}$ and then outputs $f(r)$.  
 \end{itemize}

}}
\end{tcolorbox}
\end{center}
\vspace{-5mm}
\caption{Pseudorandom function's simulator.} 
\label{fig::PRF-SIM}
\end{figure}

\begin{figure}[H]
\setlength{\fboxsep}{1pt}
\begin{center}
    \begin{tcolorbox}[enhanced,width=4.7in,left=0.1cm, 
    drop fuzzy shadow southwest,
    colframe=black,colback=white]
{\small{
%
%
%
 
  \doubleunderline{\send($C^{\sss i}, .$)}\\
 
 This query is dealt with as below: 
 
 \begin{itemize}[leftmargin=.4cm]
 \item if the client's instance is not in the ``expecting'' state and it receives query  \send($C^{\sss i}, \text{start}, \text{phase}$), where $\text{phase}\in \{\text {enrolment, authentication}\}$ then it: 
  \begin{enumerate}
  \item generates pair $(C_{\sss ID},  \text{phase})$. 
  \item responds to the query with $(C_{\sss ID},  \text{phase})$. 
  \item sets the client's instance state to expecting.
  \end{enumerate}
 \item if the client's instance state is in expecting, then: 
 \begin{itemize}[leftmargin=.4cm]
 \item upon receiving \underline{\send($C^{\sss i}, (\bar M, \bar t)$)}, it:
  \begin{enumerate}
  \item authenticates and decrypts the ciphertext $\bar M$ as $(p, b) \gets \mathsf{Dec}_{\sss\VC{k}}(\bar M, \bar t)$. If the authentication fails (i.e., $b\neq 1$), it halts.  
  \item extracts $(\VM{\nonce}, \VM{\counter})$ from plaintext $p$ and checks if $\VM{\counter} > \VC{\counter} $. If the check fails, it halts. 
  \item generates $\VC{\verifier}$ using $\VC{\salt}$ and $\VC{\pin}$ as follows $\VC{\verifier} \gets \prf_{\VC{\salt}}(\VC{\pin})$. Then, it updates its state as follows: $\forall i, 1\leq i\leq \VM{\counter}- \VC{\counter}: \text{\ (a)\ } \VC{\counter} \gets \VC{\counter} + 1\text{\ and (b) \ } (k,  \VC{\state}) \gets \update(\VC{\state}, \VC{\counter})$. 
  \item encrypts $p'=\VM{\nonce}||\VC{\verifier}$ using key $k$  as follows: $(\bar M', \bar t') \gets \enc_{\sss k}(p')$, which results in a ciphertext $\bar M'$ and tag $\bar t'$. 
  \item responds to the query with $(\bar M', \bar t')$. It sets the client's instance state to ``not expecting''.  
  \end{enumerate}
 \item  upon receiving \underline{\send($C^{\sss i}, (\ddot M', \ddot t'), (\hat M', \hat t')$)}, it: 
 \begin{enumerate}
 \item   authenticates and decrypts the ciphertext $\hat M'$ as follows: $(p', b') \gets \mathsf{Dec}_{\sss\VC{k}}(\hat M', \hat t')$. If the authentication fails (i.e., $b'\neq 1$), it halts.  
  \item extracts $(\tmp_{\sss\VM{\counter}}, \VM{\counter})$ from $p'$ and checks if $\tmp_{\sss\VM{\counter}} > \VC{\counter} $. If the check fails, it halts. 
 \item updates its state as follows: $\forall i, 1\leq i\leq \tmp_{\sss\VM{\counter}}- \VC{\counter}: \text{\ (a)\ } \VC{\counter} \gets \VC{\counter} + 1\text{\ and (b) \ } (k,  \VC{\state}) \gets \update(\VC{\state}, \VC{\counter})$. 
 \item  authenticates and decrypts $\ddot M'$ as follows: $(p, b) \gets \mathsf{Dec}_{\sss k}(\ddot M', \ddot t')$. If the authentication fails (i.e., $b\neq 1$), it halts.
 \item    extracts $(\VM{\nonce}, \VM{\trans})$ from plaintext $p$.  
 \item runs the predicate, $y \leftarrow\phi(\VM{\trans}, \pi)$. If $y=0$, it halts. 
\item generates $\VC{\verifier}$ using $\VC{\salt}$ and $\VC{\pin}$ as follows, $\VC{\verifier} \gets \prf_{\sss\VC{\salt}}(\VC{\pin})$.
 \item updates its state one more time as follows, $(k',  \VC{\state}) \gets \update(\VC{\state}, \VM{\counter})$.
 \item computes  the authenticator: $\bar{\VC{\mathit{response}}} \gets \prf_{\sss k'}( \VM{\nonce} || \VM{\trans} || \VC{\verifier}|| 1)$ and session key:  $\bar {sk}^{\sss C} \gets \prf_{\sss k'}( \VM{\nonce} || \VM{\trans} || \VC{\verifier}|| 2)$.
 \item responds the send query with  $\bar{\VC{\mathit{response}}}$. It makes the client's instance accept the key and then terminates the instance. 
 \end{enumerate}
 \end{itemize}
 \end{itemize}
}}
To keep track of all the exchanged messages, it stores the above incoming and going messages in vector $\vv L$. So, we have $\Big((C_{\sss ID},  \text{enrolment}),$ $ (C_{\sss ID},  \text{authentication}), $ $(\bar M, \bar t), $ $(\bar M', \bar t'),$ $(\ddot M', \ddot t'), $ $(\hat M', \hat t'), $ $\bar{\VC{\mathit{response}}}\Big)\in \vv L$.
\end{tcolorbox}
\end{center}
\vspace{-5mm}
\caption{Simulators for \send\ query to  a client's instance.} 
\label{fig::Send-sim-to-client}
\end{figure}

\begin{figure}[H]
\setlength{\fboxsep}{1pt}
\begin{center}
    \begin{tcolorbox}[enhanced,width=4.7in, left=0.1cm,
    drop fuzzy shadow southwest,
    colframe=black,colback=white]
{\small{

 \doubleunderline{\send($S^{\sss j}, .$)}\\
 
 This query is dealt with as below: 
 
 \begin{itemize}[leftmargin=.4cm]
 \item upon receiving  \underline{\send($S^{\sss j}, (C_{\sss ID},  \text{enrolment})$)}, it:
 \begin{enumerate}
 \item  increments its counter as $\VS{\counter} \gets \VS{\counter} + 1$, updates its state as $\bar{\VS{\keyt}_{\sss 1}}, \VS{\state} \gets \update(\VS{\state}, \VS{\counter})$, picks a random value $\bar{\VS{\nonce}} \stackrel{\sss\$}\leftarrow \bin^{\sss\secpar}$, and generates ciphertext and tag   $(\bar M,\bar t) \gets \enc_{\sss\VS{k}}(\bar{\VS{\nonce}}|| \VS{\counter})$. 
 
 \item responds to the query with $(\bar M,\bar t)$. The state of the server instance is set to ``expecting''.
 %
 %
 \end{enumerate}
 \item upon receiving  \underline{\send($S^{\sss j}, (\bar M',  \bar t'$))}, it:
 \begin{enumerate}
 \item authenticates and decrypts the ciphertext $\bar M'$ as $(p', b') \gets \mathsf{Dec}_{\sss\bar{\VS{\keyt}_{\sss 1}}}(\bar M', \bar t')$. If the authentication fails (i.e., $b'\neq 1$), it halts. 
\item    extracts $(\VM{\nonce}, \VM{\verifier})$ from plaintext $p'$. It sets $\VS{\verifier}\gets\VM{\verifier}$ and also  checks if $\VM{\nonce}=\bar{\VS{\nonce}}$. If the equation does not hold, it halts. The state of the server instance is set to expecting.
 \end{enumerate}
 \item upon receiving  \underline{\send($S^{\sss j}, (C_{\sss ID},  \text{authentication})$)}, it:
 \begin{enumerate}
  \item  increments its counter  $\VS{\counter} \gets \VS{\counter} + 1$, updates its state  $\bar{\VS{\keyt}_{\sss 2}}, \VS{\state} \gets \update(\VS{\state}, \VS{\counter})$,  temporarily stores this counter $\tmp_{\sss\VS{\counter}} \gets \VS{\counter}$, increments the counter again $\VS{\counter} \gets \VS{\counter} + 1$, updates its state again $\bar{\VS{\keyt}_{\sss 3}}, \VS{\state} \gets \update(\VS{\state}, \VS{\counter})$, and picks a random value $\ddot{\VS{\nonce}} \stackrel{\sss\$}\leftarrow \bin^{\sss\secpar}$.
  \item generates two pairs of ciphertext and tag as follows, $(\ddot M', \ddot t')\leftarrow \enc_{\sss\bar{\VS{\keyt_{\sss 2}}}}(\ddot{\VS{\nonce}} || \VS{\trans})$ and  $(\hat M', \hat t')\leftarrow \enc_{\sss\VS{k}}(\tmp_{\sss\VS{\counter}} || \VS{\counter})$. 
  \item responds to the query with $(\ddot M', \ddot t')$, $(\hat M', \hat t')$. The state of the server instance is set to expecting.
 \end{enumerate}
 \item upon receiving  \underline{\send($S^{\sss j},   \bar{\VC{\mathit{response}}}$)}, it:
 \begin{enumerate}
 \item computes $\VS{\mathit{expected}} \gets \prf_{\sss\bar{\VS{\keyt_{\sss3}}}}(\ddot{\VS{\nonce}}  \| \VS{\trans} \| \VS{\verifier}|| 1)$. It checks whether  $ \bar{\VC{\mathit{response}}}=\VS{\mathit{expected}}$. If the equality does not hold, the server instance terminates without accepting any session key. 
  \item generates the session key $\bar {sk}^{\sss S} \gets \prf_{\sss\bar{\VS{\keyt_{\sss3}}}}(\ddot{\VS{\nonce}}  \| \VS{\trans} \| \VS{\verifier}|| 2)$. 
It accepts the key and terminates.
  \end{enumerate}
 \end{itemize} 
}}
\end{tcolorbox}
\end{center}
\caption{Simulators for \send\ query to  a server's instance.} 
\label{fig::Send-sim-to-server}
\end{figure}

\begin{figure}[H]
\setlength{\fboxsep}{1pt}
\begin{center}
    \begin{tcolorbox}[enhanced,width=4.7in, left=0.1cm, 
    drop fuzzy shadow southwest,
    colframe=black,colback=white]
{\small{

 \doubleunderline{\execute($C^{\sss i}, S^{\sss j}$)}\\
 
 This query is dealt with as below: 
 \begin{enumerate} 
 \item $(C_{\sss ID},  \text{enrolment})\gets$\send($C^{\sss i}, \text{start}, \text{enrolment}$).
\item  $(\bar M,\bar t)\gets$\send($S^{\sss j}, (C_{\sss ID},  \text{enrolment})$).
\item $(\bar M', \bar t')\gets$\send($C^{\sss i}, (\bar M, \bar t)$).
\item $(C_{\sss ID},  \text{authentication})\gets$\send($C^{\sss i}, \text{start}, \text{authentication}$).
\item $(\ddot M', \ddot t', \hat M', \hat t')\gets$\send($S^{\sss j}, (C_{\sss ID},  \text{authentication})$).
\item $\bar{\VC{\mathit{response}}}\gets$\send($C^{\sss i}, (\ddot M', \ddot t'), (\hat M', \hat t')$).
  \item outputs the following transcript: $[(C_{\sss ID},  \text{enrolment}), (\bar M, \bar t), (\bar M', \bar t'),(C_{\sss ID},  $ $\text{authentication}), $ $(\ddot M', \ddot t'), (\hat M', \hat t'), \bar{\VC{\mathit{response}}}]$.
 \end{enumerate} 
  \noindent\rule{4.6in}{1pt}
  \doubleunderline{\reveal($I$)}\\
  
This query is processed as follows.
  
\ $\bullet$ returns session key $\bar {sk}^{\sss I}$ (computed by $I\in\{C,S \}$), if $I$ has already accepted the key.

  \noindent\rule{4.6in}{1pt}
    \doubleunderline{\test($I$)}\\
    
  This query is processed as below.  
  
 \begin{enumerate} 
 \item $sk\gets$\reveal($I$).
 \item $b\stackrel{\sss\$}\gets\{0,1\}$.
 \item sets $v$ as follows:  \begin{equation*}
   v= 
\begin{cases}
    sk,              &\text{if } b= 1\\
   r \stackrel{\sss\$}\gets\{0,1\}^{\sss \iota}, & \text{otherwise }\\
\end{cases}
\end{equation*}
 \item returns $v$.
 \end{enumerate}
}}
\end{tcolorbox}
\end{center}
\caption{Simulators for \execute, \reveal, and \test\ queries. } 
\label{fig::sim-for-exe-rev-test}
\end{figure}

\subsection{Authentication}

In this section, we prove the protocol's authentication. We begin with the case where the adversary $\adv$ has access to the traffic between the two parties and wants to impersonate the client, $C$; we denote such a case with $\bar {aut}$. The Authentication proof relies on the semantic security proof (and games) we presented in Section \ref{sec::semSec-proof}.  Now, we outline the proof.  By definition, it holds that:
 \begin{equation}\label{eq::adv-ake}
Adv_{\sss \psi}^{\sss \bar {aut}}(\A)=Pr[Auth_{\sss  0}]
\end{equation}

Also, we can extend Equation \ref{eq::game_1} to:
\begin{equation*}
|Pr[Auth_{\sss  1}]-Pr[Auth_{\sss  0}]| \leq (q_{\sss s}+q_{\sss p})Adv^{\sss\prf}(\adv),
\end{equation*}
 because the only difference between the two games (i.e., $G_{\sss  0}$ and $G_{\sss  1}$) is that the output of the $\prf$ is replaced with an output of a uniformly random function $f$.

 Furthermore, we can extend Equations \ref{eq::game_1}-\ref{eq::game_7} as follows:

 \begin{equation*}
 \begin{split}
  &  |Pr[Auth_{\sss  1}]-Pr[Auth_{\sss  0}]| \leq (q_{\sss s}+q_{\sss p}) Adv^{\sss \prf}(\adv) \\ 
 &  |Pr[Auth_{\sss  2}]-Pr[Auth_{\sss 1}]| \leq (q_{\sss s}+q_{\sss p}) Adv^{\sss Enc}(\adv) \\ 
 & Pr[Auth_{\sss  3}]-Pr[Auth_{\sss  2}]=0\\
 & |Pr[Auth_{\sss  4}]-Pr[Auth_{\sss  3}]|\leq \frac{4(q_{\sss  s}+q_{\sss  p})}{2^{\sss \lambda}}\\
 & |Pr[Auth_{\sss  5}]-Pr[Auth_{\sss  4}]|\leq \frac{q_{\sss  s}}{2^{\sss \lambda}}\\
 & |Pr[Auth_{\sss  6}]-Pr[Auth_{\sss  5}]|\leq \frac{q_{\sss  s}}{2^{\sss \lambda}}\\
 & |Pr[Auth_{\sss  7}]-Pr[Auth_{\sss  6}]|\leq \frac{2q_{\sss  s}}{2^{\sss \lambda}}
\end{split}
\end{equation*}

 Moreover, since the authenticator is a random value in $G_{\sss 7}$, it holds that $Pr[Auth_{\sss 7}]=\frac{q_{\sss  s}}{2^{\sss \lambda}}$. We conclude the proof, by summing up the above relations and combining with Equation \ref{eq::adv-ake}: 
  \begin{equation}
 Adv_{\sss \psi}^{\sss \bar {aut}}(\A) = Pr[Auth_{\sss  0}] \leq (q_{\sss s} + q_{\sss p})\Big(Adv^{\sss\prf}(\adv)+Adv^{\sss Enc}(\adv)\Big)+\frac{9q_{\sss  s}+4q_{\sss  p}}{2^{\sss \lambda}}
 \end{equation}

Next, we proceed to the case where the adversary is given further access to the PIN, i.e., $\adv$ can also send query  \corrupt($C, 1$). We argue that given such an extra capability does not affect the adversary's advantage and the above analysis (as the protocol and its analysis have relied on the security of the CCA-secure symmetric encryption and $\prf$). Now move on to the case where $\adv$ (a) is given all the parameters stored in the hardware token, and (b) has access to all the traffic between the two parties, i.e., $\adv$ can also send query  $\bar{\text{Cpt}_{\sss  2}} =$ \corrupt$(C, 2)$. We argue that in this case, the upper bound of $\adv$'s advantage will be changed as follows:  $Adv_{\sss\psi, \bar{\text{Cpt}_{\sss  2}}}^{\sss  aut}(\A)\leq  \cfrac{q_{\sss  s}}{N}$. The reason for such a big change is that in this case, $\adv$ has all secret parameters, except the PIN and verifier $\VC{\verifier}$. \footnote{The case where $\adv$ has the additional capability to send query \corrupt($C, 2$) was never discussed and analysed in \cite{BressonCP03}. However, we noticed that $\adv$ in that scheme would have the same upper bound advantage as $\adv$ in our scheme does.} Thus, when we take the forward security into account, the advantage of the adversary (due to the union bound)  is as follows: 
  \begin{equation*}
 Adv_{\sss \psi}^{\sss  aut}(\A)  \leq (q_{\sss s} + q_{\sss p})\Big(Adv^{\sss\prf}(\adv)+Adv^{\sss Enc}(\adv)\Big)+\frac{9q_{\sss s}+4q_{\sss  p}}{2^{\sss \lambda}}+  \cfrac{q_{\sss  s}}{N}
 \end{equation*}

\subsection{PIN's Privacy Against A Corrupt Server} 

In the case where the adversary (i) has access to the parties' traffic and (ii) can make query \corrupt($S, 1$), to extract all parameters of the server, then the probability that the adversary can find the valid PIN depends on the probability of finding the correct PIN and finding $C$'s correct key of $\prf$; therefore, the probability is at most $\cfrac{q_{\sss  p}}{2^{\sss \lambda}N}$.


\section{Related Work}

In this section, first, we briefly discuss the common approaches for generating a One-Time Password (OTP) which yields from a combination of a PIN and a hardware token.  After that, we provide an overview of hardware token variants.

\subsection{Common Approaches for Generating OTP}

In the authentication that relies on a combination of knowledge and possession factors, once the client enters the secret into the hardware token, the device (in some cases after validating the secret) combines this secret with the output of one of the following methods to generate a unique OTP:  (i) a random challenge: this approach requires the server to send a random challenge to the device (through the client); those protocols that use this approach needs to ensure the random challenges themselves remain confidential in the presence of an eavesdropping adversary,  (ii) an internal counter:  the solutions that use this approach needs to take into account the situation where the token-side counter becomes out of synchronisation, or (iii) the current accurate time: this approach requires the authentication server and token use a synchronised clock and the two endpoints may get out of synchronisation after a certain time. There exist 2FA solutions (including the one we propose in this paper) that employ a combination of the above approaches. 
\subsection{Variants of OTP Hardware Tokens}

\subsubsection{Connected Tokens.}
This type of token requires a client to physically connect the token to their computer (e.g., a laptop or card reader) via which the client is authenticating. Once it is connected, the device transmits the authentication information to the computer (either automatically or after pressing a button on the token). USB tokens and smart cards are two popular token technologies in this category.  Various companies including Google, Dropbox, and  ``Fast IDentity Online'' (FIDO) Alliance have developed USB hardware tokens. YubiKey\footnote{https://www.yubico.com} is one of the well-known ones developed by FIDO. The FIDO  Alliance has proposed a standard that aims at allowing clients to log in to remote services with a local and trusted authenticator. It supports a wide range of authentication technologies including USB (security) tokens. However, researchers have discovered various vulnerabilities within this standard via manual, e.g., in \cite{PanosMNPX17,ChangMSS17,LoutfiJ15} and formal analysis, e.g., in \cite{ndss/FengLP021}. 

Smart card technology is another authentication means which has been widely used. Often it comprises two separate components; namely, a smart card and a card reader, where the former includes an integrated secure chipset while the former includes a keypad and a screen. Since its introduction in \cite{chang1991remote}, there have been numerous protocols for smart card-based 2FA (e.g., in \cite{gupta2021machine,WangW18,radhakrishnan2022dependable}) along with a few works that identify vulnerabilities of existing solutions, e.g., in \cite{TianLHL20,WangGCW16,ChaturvediDMM16}. However, the existing smart card-based solutions (e.g., in \cite{gupta2021machine,WangW18,radhakrishnan2022dependable}) are often based on public-key cryptography which imposes a high computation cost and makes the card reader's battery run out relatively fast; also some solutions (e.g., in \cite{kim2009more}) rely on tamper-proof secure chipsets embedded in the card which would ultimately increase the device's cost.


\subsubsection{Disconnected Tokens.}

This type of token does not have a physical connection to a client's computer making them more convenient than connected tokens. A disconnected token is often equipped with a built-in screen and a keypad letting a client type in the knowledge factor and view the OTP on the screen (see below for an exception).  Below, we provide an overview of two main categories of disconnected tokens.

\begin{enumerate}
\item \underline{Dedicated hardware-based Tokens}, such as RSA SecureID \cite{secureID}, OneSpan Digipass 770 \cite{Digipass-website}, and Thales Gemalto SWYS QR Token Eco \cite{Gemalto}.   RSA SecureID (unlike the other two tokens) does not have a keypad. Briefly, in RSA SecureID, the OTP is generated using the current time and a secret key (allocated to the client and) stored in the token \cite{biryukov2003cryptanalysis}. Thus, not only the token has to have a synchronised clock with the server, but also the token's OTP can be generated by an adversary who has physical access to the device, as it can extract the device's secret key.  The main advantage of  Digipass 770 and Thales Gemalto SWYS QR Token Eco to RSA SecureID is that they allow clients to see and verify the transaction details through the token. Therefore, the client is given more understandable information about the transaction it is approving,
so phishing (by Man-in-the-Browser attacks or social engineering attacks) becomes harder.

Our investigation suggests that Digipass 770 also \emph{locally stores and verifies} clients' PINs. 
%
Specifically, once a client receives the token, it also receives an activation code from the verifier, e.g., the client's bank.  Then, the client (i) registers the activation code in the device and (ii) registers the activation code to the verifier, so the verifier knows that this specific client has a device with the provided activation code. Then, the client registers its PIN in the device which stores it locally. Every time a client uses the verifier's online system  (e.g., online banking) and makes a transaction, the system generates and displays an encrypted visual image. The client uses its token (camera) to scan the image, and then enters its PIN into the device. Next, the device checks the PIN; if the PIN matches the previously registered PIN, then it decrypts the image and displays the transaction's content on the token's screen which allows the client to check whether the transaction is the one it has made. If the client accepts the transaction and presses a certain button, then the token generates and displays an OTP that the client can insert into the verifier's online system \cite{Digipass-website,DIGIPASS-doc}.  Thales Gemalto SWYS QR Token Eco also uses a mechanism similar to the one we described above. 


Jules et al. \cite{juels2016configurable} discussed that the adversary who can intercept the client and server's communication and also has physical access to the client's token or the server's storage can extract the client's PIN and impersonate the client. To address the issue they also suggested a solution that can address the above issue by using (i) a forward-secure pseudorandom number generation, (ii) multiple servers, etc. However, the proposed scheme lacks formal proof and does not consider the case where transactions' details must be verified by clients on the token.  

Moreover, Jarecki \textit{et al.} \cite{JareckiJKSS21} proposed a (single server) protocol to ensure that even if the server or the device is corrupted a client's PIN cannot be extracted and the adversary cannot impersonate an honest client. It is mainly based on a hash function, both symmetric and asymmetric key encryptions, and (Diffie–Hellman) key exchange. This scheme has a high computation and communication cost due to its complexity,  the use of public-key cryptography, and numerous rounds of communication, even between the client and token. Also, it requires the token to perform asymmetric-key operations and invoke symmetric key primitives many times, which would make the token's battery run out quickly. This protocol requires the client (in addition to remembering its PIN) to remember/store a cryptographic secret key locally (but not on the token), as a result of invoking a subroutine called asymmetric  ``password-authenticated key exchange'' (PAKE). Furthermore, there is another authentication protocol, that does not rely on a trusted chipset, presented in \cite{zhang2020strong}. Nevertheless, it has been designed for ``federated identity systems'' and is not suitable for two/multi-factor authentication settings.


%
\item \underline{Mobile phone-based Tokens}, such as the solutions presented in \cite{SARA22,KoganMB17,KonothFFARB20}. There have been protocols that generate an OTP with the use of a mobile phone as a hardware token. Such solutions often rely on the added features that mobile phones offer, such as possessing a Trusted Execution Environment (TEE), being able to communicate directly with the server, or having a rechargeable battery. The scheme in \cite{KoganMB17} relies on a combination of time-based OTP and a hash chain. This scheme ensures that even if the adversary corrupts the server at some point, then it cannot extract the client's secret. Nevertheless, it  (a) requires the client to store a sufficiently long secret key (on the mobile phone), (b) requires the laptop/PC that the client uses to be equipped with a camera, and (c) needs the mobile phone to invoke a hash function over a million times that can cause the phone's battery to run out fast. The protocol proposed in \cite{KonothFFARB20} mainly relies on a phone's TEE (i.e.,  ARM TrustZone technology) and messages that the server can directly send to the phone. Later,  Imran \textit{et al.} \cite{SARA22} proposes a new protocol that also relies on a phone's TEE, but it improves the protocol presented in \cite{KonothFFARB20}, in the sense that it is compatible with more android devices and supports biometric authentication too. 

A primary limitation of mobile phone-based OTP tokens is that they cannot be used when there is no (mobile phone) network coverage. Another limitation is that in certain cases (beyond internet banking) sharing phone numbers with the authentication server may not suit all clients, e.g., transactions' details along with the phone number might be sold for targeted advertisements. 
\end{enumerate}

%





\section*{Acknowledgements}

Steven J. Murdoch and Aydin Abadi were supported by REPHRAIN: The National Research Centre on Privacy, Harm Reduction and Adversarial Influence Online, under UKRI grant: EP/V011189/1. Steven J. Murdoch was also supported by The Royal Society under grant UF160505. 

\bibliographystyle{splncs03}
\bibliography{ref}
\appendix

\section{Definition and Construction of Forward-Secure Pseudorandom Bit Generator}\label{sec::def-FS-PRG}

In this section, we restate the formal definition of the forward-secure pseudorandom bit generator (taken from \cite{BellareY03}), and then briefly explain how it can be constructed. A standard pseudorandom generator is said to be secure if its output is computationally indistinguishable from a random string of the same length. However, the forward security of a stateful generator requires more security guarantees. Specifically, in this setting,  an adversary $\mathcal{A}$ may at some point penetrate the machine in which the state is stored and obtain the current state. In this case, the adversary is able to compute the future output of the generator. But, it is required that the bit strings generated in the past still be secure, i.e., the strings are computationally indistinguishable from random bit strings. This implies that it is computationally infeasible for the adversary to recover the previous state from the current one.

In this setting, the adversary is allowed to choose when it wants to penetrate the machine, as a function of the output blocks it has seen so far. Thus, first, the adversary runs in a ``find'' stage where it is fed output blocks, one at a time, until it says it wants to break in, and at that time the current state is returned.  Next, in the ``guess'' stage, it must decide if the output blocks that were given to it were the outputs of the generator, or were independent random bits. This is captured formally by two experiments; namely, real and random. In the real experiment, the forward secure generator is used to generate output blocks. Nevertheless, in the ideal experiment, the output blocks are truly random strings (of the same length as that of the blocks in the real experiment). Note that below ``$\mathcal{A}(\text{find}, out, h)$'' denotes $\mathcal{A}$ in the find stage, and is given an output block $out$ and current history $h$ and returns a pair $(I, h)$ where $h$ is an updated history and $I \in\{\text{find}, \text{guess}\}$. Below, we restate the two experiments.

\

\begin{minipage}{57mm}
\begin{tcolorbox}[left=0mm]
$
  \begin{array}{l}
\underline{\mathsf{Exp}_{\sss \text{real}}^{\sss \text{fs-prg}}(\mathcal{A},\text{aux})}\\
\\
{st}_{\sss 0}\stackrel{\sss \$}\leftarrow \mathsf{FS\text{-}RPG}.\kgen(1^{\sss \lambda}) \\
i\gets 0\\ h\gets \text{aux} \\
\text {Repeat}\\
i\gets i+1\\
\hspace{4mm} ({out}_{\sss i}, st_{\sss i})\leftarrow    \mathsf{FS\text{-}RPG.next}    (st_{\sss i-1})\\
\hspace{4mm}  (I, h)\gets\mathcal{A}(\text{find}, out_{\sss i}, h)\\
\text{Until}\hspace{2mm}  (I=guess)\hspace{2mm}  \text{or} \hspace{2mm}  (i=n)\\
g\gets\mathcal{A}(\text{guess}, st_{\sss i}, h)\\
\text{Return} g\\
  \end{array}
  \qquad$
  \end{tcolorbox}
   \end{minipage}
   \begin{minipage}{57mm}
\begin{tcolorbox}[left=0mm]
$
  \begin{array}{l}
 \underline{\mathsf{Exp}_{\sss \text{ideal}}^{\sss \text{fs-prg}}(\mathcal{A}, \text{aux})}\\
  \\
   {st}_{\sss 0}\stackrel{\sss \$}\leftarrow \mathsf{FS\text{-}RPG}.\kgen (1^{\sss \lambda}) \\
i\gets 0; h\gets \text{aux} \\
\text {Repeat}\\
i\gets i+1\\
\hspace{4mm} ({out}_{\sss i}, st_{\sss i})\leftarrow  \mathsf{FS\text{-}RPG.next}   (st_{\sss i-1})\\
\hspace{4mm}{out}_{\sss i}\stackrel{\sss \$}\leftarrow\{0, 1\}\\
\hspace{4mm}  (I,h)\gets\mathcal{A}(\text{find}, out_{\sss i}, h)\\
\text{Until}\hspace{2mm}  (I=guess)\hspace{2mm}  \text{or}\hspace{2mm}  (i=n)\\
g\gets\mathcal{A}(\text{guess}, st_{\sss i}, h)\\
\text{Return} g\\
  \end{array}
$
\end{tcolorbox}
   \end{minipage}
   
\

Given the experiments, the adversary's advantages are defined in the following two equations.

\begin{equation}\label{equ::adv-1st-term}
\mathtt{Adv}^{\sss \text{fs-prg}}(\mathcal{A})=Pr[\mathsf{Exp}_{\sss \text{real}}^{\sss \text{fs-prg}}(\mathcal{A},\text{aux})=1]-Pr[\mathsf{Exp}_{\sss \text{ideal}}^{\sss \text{fs-prg}}(\mathcal{A},\text{aux})=1]
\end{equation}

\begin{equation}\label{equ::adv-2nd-term}
\mathtt{Adv}^{\sss \text{fs-prg}}(t)= Max\{\mathtt{Adv}^{\sss \text{fs-prg}}(\mathcal{A})\}
\end{equation}

Equation \ref{equ::adv-1st-term}  refers to the (fs-prg) advantage of $\mathcal{A}$ in attacking the forward-secure pseudorandom bit generator, FS-PRG. While Equation \ref{equ::adv-2nd-term} refers to the maximum advantage of $\mathcal{A}$ in attacking FS-PRG, where the adversary has a time-complexity at most $t$. It is required that the adversary's advantage is negligible for practical values of $t$.

Bellare \textit{et al.} \cite{BellareY03} proposed various instantiations of FS-PRG, including the one based on AES. In the latter case, one can set a block size $b$ and a state size $s$ to $128$ bits. We refer readers to \cite{BellareY03} for further discussion. 


\end{document}